\documentstyle[12pt,a4,epsf]{article}
\newcommand{\D}{\partial}
\newcommand{\be}{\begin{equation}}
\newcommand{\ee}{\end{equation}}
\newcommand{\Bx}{{\bf x}}

\newcommand{\Bq}{{\bf q}}
\newcommand{\Bp}{{\bf p}}
\newcommand{\TR}{^{\mbox{\tiny T}}}
\newcommand{\F}{\frac}
\newcommand{\FD}[1]{\F{\delta}{\delta \phi\left(#1\right)}}
\newcommand{\FDM}[1]{\F{\bar\delta}{\bar\delta \phi\left(#1\right)}}
\def\JLone<#1,#2>{#1}
\def\JLtwo<#1,#2,#3>{#2}
\def\JLyear<#1,#2,#3,#4>{#3}
\def\JLpage<#1,#2,#3,#4>{#4}
\def\JL#1{\JLone<#1>\ {\bf \JLtwo<#1>},  \JLpage<#1> (\JLyear<#1>)}
\def\Jpage<#1,#2,#3>{#3}
\def\andvol#1{{\bf \JLone<#1>}, \Jpage<#1> (\JLtwo<#1>)}
\def\PTP#1{Prog.\ Theor.\ Phys.\ \andvol{#1}}

\def\PR#1{Phys.\ Rev.\ \andvol{#1}}
\def\PRL#1{Phys.\ Rev.\ Lett.\ \andvol{#1}}
\def\PL#1{Phys.\ Lett.\ \andvol{#1}}
\def\NP#1{Nucl.\ Phys.\ \andvol{#1}}

\def\IJMP#1{Int.\ J.~Mod.\ Phys.\ \andvol{#1}}

\begin{document}                
\begin{flushright}
\parbox{4cm}{
KUCP-0147\\
KANAZAWA-99-25}
\end{flushright}
\begin{center}
{\large\bf
Scheme Dependence of the Wilsonian Effective Action\\
and Sharp Cutoff Limit of the Flow Equation}
\end{center}
\vskip4mm
\begin{center}
{Jun-Ichi Sumi, Wataru Souma, Ken-Ichi Aoki$^*$, 
Haruhiko Terao$^*$\\ and Keiichi Morikawa$^{**}$}
\end{center}
\vskip2mm
\begin{center}
{\it
Department of Fundamental Sciences,
 Faculty of Integrated Human Studies,\\
 Kyoto University, Kyoto 606-8501, Japan\\
$^*$Institute for theoretical Physics, Kanazawa University, \\
Kakuma-machi, Kanazawa 920-1192, Japan\\
$^{**}$Research Center for Nanodevices and Systems,
Hiroshima University,\\
1-4-2 Kagamiyama, Higashi-Hiroshima 739-8527, Japan}
\end{center}

\begin{abstract}
The cutoff scheme dependence in the several formulations of the Exact 
Renormalization Group (ERG) is investigated.
It is shown that the cutoff scheme dependence of the 
Wilsonian effective action is regarded as a certain coordinate 
transformation on the theory space. 
From this observation the Wilsonian effective actions  are found to 
suffer from strong dependence on the schemes even in the infra-red 
asymptotic region for massive theories. 
However there is no such scheme dependence in the one particle irreducible 
parts of them, which is called the effective average actions.
We also derive the explicit form of the Polchinski RG equation in the sharp
cutoff limit. 
Finally this equation is shown to be identical with the 
Wegner-Houghton RG equation.
\end{abstract}

\section{Introduction}               

The Exact renormalization group (ERG) \cite{rf:RGE} has been 
one of the analytical tools to investigate non-perturbative 
phenomena of field theories, 
(e.g. the chiral symmetry breaking \cite{rf:chiral} etc.).
The ERG flow equations are the functional differential equations for the 
Wilsonian effective actions $S_\Lambda[\phi]$, where $\Lambda$
is an ultra-violet (infra-red) momentum cutoff of the low energy modes
$\phi(\Bq)$ (the high energy modes which are already integrated out). 
The responce of the effective action under variation of the cutoff is 
exactly represented as 
\be
\F{\D}{\D\Lambda}S_\Lambda[\phi]=F[S_\Lambda],
\ee
where $F[S_\Lambda]$ is a finite functional of the field $\phi$.
The explicit forms of $F[S_\Lambda]$ are shown later. 
By solving the ERG flow equations toward to $\Lambda=0$ with  
certain bare actions as the initial conditions, we can obtain 
the generating functionals of the connected Green's functions.
There have been also known another type of the ERG equations for the 
cutoff Legendre effective action (or the effective average action) 
$\Gamma_{\Lambda}[\phi]$.
In this case the solutions of the ERG lead to the ordinary effective 
actions, or the generating functionals of the 1PI Green's functions.

Usually, we write down the ERG equations for the dimensionless parameters
in the effective actions by scaling the parameters with the infra-red 
cutoff $\Lambda$, 
since the energy unit used to represent the theories does not have any 
physical significance. 
The functional space of the dimensionless effective action is called the 
theory space. 
Through this manipulation the beta-functional $F$ becomes free from the
scale $\Lambda$.
Among the RG flows of these dimensionless quantities, especially the fixed
points, the critical surfacies and the renormalized trajectories, 
are of utility indispensable in investigating (statistical) continuum limit
of field theories. 

It is important for the practical analyses that the ERG admits 
non-perturbative as well as systematic approximations;
e.g. the derivative expansion \cite{rf:MOR2}, 
the momentum scale expansion \cite{rf:MOR1,rf:MOR4}. 
Though we used the word of `expansion' here, the approximation schemes are 
not the series expansions with respect to some explicit small parameters. 
This is an essential distinction from the ordinary expansion schemes; 
$\epsilon$-expansion, $1/N$-expansion and perturbation, which lead to
the asymptotic series. 
The solutions of the ERG equations are expected to converge smoothly 
with the improvement of approximatioins. 
Furthermore, we may obtain fairly good non-perturbative results 
within the simple approximation schemes \cite{rf:SO}.

\begin{figure}[htb]
\epsfxsize=0.5\textwidth
\centerline{\epsfbox{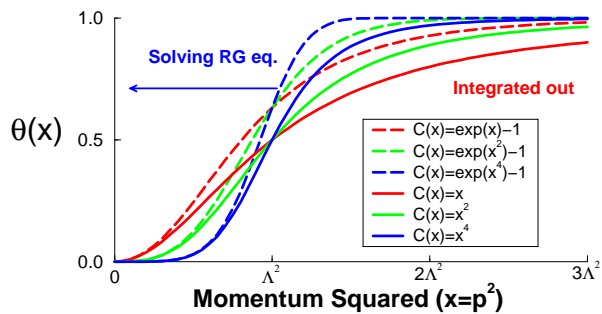}}
\caption{The examples of the infra-red cutoff functions
$\theta(x)=xC(x)/(1+xC(x))$ for various $C(x)$. 
$C^{-1}(x)$ is the mass function introduced in
Eq.\ (\ref{eq:cutofffun}).}
\label{fig:1}
\end{figure}

The ERG flow equation depends on the cutoff schemes. Here the cutoff
scheme means the profile of the cutoff function in the propagator. 
It is convenient to perform the cutoff of the infra-red region 
$p^2<\Lambda^2$ by adding a momentum dependent mass, 
\be
\Delta S[\phi]\equiv
\int d^dx \F{\Lambda^2}{2}\phi\cdot
 C^{-1}(-\D^2/\Lambda^2) \cdot\phi,
\label{eq:cutofffun}
\ee
where $C$ is a proper cutoff function satisfiying that 
$C(x)\to 0$ as $x\to 0$ and $d$ is the space-time dimensions. 
Then the propagator is multipied by the cutoff function 
$\theta(x)=xC(x)/(1+xC(x))$.
In Fig.\ \ref{fig:1} the examples of the cutoff functions for various
$C(x)$ are shown. The sharp cutoff scheme corresponds to the step function; 
$\theta(p^2)=0$ for $p<\Lambda$ and $\theta(p^2)=1$ for $p>\Lambda$. 

The effective actions treated by the ERG equations themselves are 
cutoff scheme dependent even after the cutoff is removed; 
$\Lambda \rightarrow 0$.
While the physical quantities obtained from their continuum limit, 
or the renormalized trjectories, should not be affected by the 
regularization scheme. 
Therefore it will be important to see the cutoff scheme dependence of 
the RG flows, specially the renormalizaed trajectories, and to find out
the scheme independent quantities at $\Lambda=0$ which are the 
physical quantities obtained in the ERG approach.
In this paper we discuss the basic strucure of the cutoff scheme
dependence of the ERG equations and of thier solutions; the effective
actions. 
Especially we look into the scheme dependence of the low energy effective 
actions, or the renormalized trajectories, in the asymptotic regions for 
massive theories. 
In such regions the Wilsonian effectives actions are found to suffer from 
strong scheme dependence, while the Legendre effective actions are
free from such problems.
Related with this we also examine the sharp cutoff limit of the
so-called Polchinksi equation by taking care of the singular limit of the
cutoff profiles.
Moreover it will be shown that the Polchinksi equations turn out to be
equivalent to the Wegner-Houghton equation in the sharp cutoff limit.
This equivalence between these two formulations of the ERG has not been
proved yet.

This paper is organized as follows. 
In Sec.\ \ref{sec:2} we briefly overview the derivation of the ERG flow 
equations; the Wegner-Houghton equation, the Polchinski equation and 
the flow equation for the cutoff Legendre effective action.
In Sec.\ \ref{sec:3} we will examine the general aspects of 
the scheme dependence of the effective actions. 
At first it will be shown that the variation of the cutoff function;
$\theta(x)\to\theta(x)+\delta\theta(x)$, can 
be reinterpreted as the coordinate transformation on the theory 
space.
From this observation it immediately follows that the critical exponents 
obtained by the ERG method are independent of the cutoff scheme. 
We also discuss the formulations using the cutoff mass functions depending
on the wave function renormalization from this point of view.
We will study also the cutoff scheme dependence of the renormalized 
trajectories in the infra-red asymptotic region of massive theories.
The scheme dependence of the coefficient 
function $V_k^i(\phi)$ of the Wilsonian effective action $S_\Lambda[\phi]$ 
behaves as $1/\Lambda^{2k}$ and, therefore, becomes so strong as to prevent
from taking any physical information as the infra-red cutoff $\Lambda$ is 
lowered. 
While the cutoff Legendre effective actions are free from such a 
strong scheme dependence.

The sharp cutoff limit of the Polchinski equations and their equivalence 
to the Wegner-Houghton formulation will be clarified in Sec.\ \ref{sec:4} 
and in Sec.\ \ref{sec:5}. 
Section \ref{sec:6} is devoted to the conclusions and some remarks. 
Throughout this paper, we restrict ourselves to the single 
scalar theories. 
This restriction does not loose the generality of the discussions. 

\section{ERG Equations}
\label{sec:2}

To fix our conventions and to make this paper self-contained, 
we briefly overview the derivations of the ERG flow equations. 
Let us start from the generator of the connected Green's functions 
$W[J]$ given by
\be
\exp\left(W[J]\right)=
\int D\phi\exp\left(-\Delta S_{\mbox{\tiny U.V.}}
-S_{\Lambda_0}+J\cdot\phi\right),\label{eq:A}
\ee
where $\Delta S_{\mbox{\tiny U.V.}}$ is the ultra-violet cutoff term 
regularizing the path-integral (\ref{eq:A}). 
We sometimes use the shorthand: $J\cdot\phi=\int d^dx
J(x)\phi(x)$ etc., where $d$ is the (euclidean) space-time dimensions. 
Now, we introduce the intermediate scale $\Lambda<\Lambda_0$ and 
formally integrate out the high energy modes $\phi_>(\Bq)$ 
($\Lambda<q\le\Lambda_0$). 

Then we get the effective action for the 
low energy modes $\phi(\Bq)$ ($q<\Lambda$),
\be
\exp\left(-S_\Lambda[\phi,J]\right)=
\int D\phi_>\exp\left(-\Delta S_\Lambda^{\Lambda_0}[\phi_>]
-S_{\Lambda_0}[\phi+\phi_>]+J\cdot(\phi+\phi_>)\right),
\label{eq:B}
\ee
where the cutoff action $\Delta S_\Lambda^{\Lambda_0}[\phi_>]$ is 
given by,
\be
\Delta S_\Lambda^{\Lambda_0}[\phi]\equiv
\F{1}{2}\phi\cdot \left[P_\Lambda^{\Lambda_0} \right]^{-1}
\cdot\phi.
\ee
The support of $P_\Lambda^{\Lambda_0}(q)$ is effectively restricted 
in the region $\Lambda<q<\Lambda_0$ by means of a certain smooth 
cutoff function. Furthermore, we set, 
$\Delta S_{\mbox{\tiny U.V.}}=\Delta S_0^{\Lambda_0}$ 
and $P_0^{\Lambda_0}(q)=P_0^\Lambda(q)+P_\Lambda^{\Lambda_0}(q)$. 
This can be achieved by multiplying the partition of unity 
$\theta_\Lambda^{\Lambda_0}(q)$ to the propagator $1/q^2$, 
i.e. $P_\Lambda^{\Lambda_0}(q)=\theta_\Lambda^{\Lambda_0}(q)
/q^2$. $\theta_\Lambda^{\Lambda_0}(q)$ approximately vanishes 
when $q\gg\Lambda_0$ or $q\ll\Lambda$, and 
$\theta_\Lambda^{\Lambda_0}(q)\approx 1$ for $\Lambda<q<\Lambda_0$. 
In the RG flow equation, we can safely forget the ultra-violet 
cutoff $\Lambda_0$ by taking the limit $\Lambda_0\to\infty$, since 
$\D P_\Lambda^{\Lambda_0=\infty}(q)/\D\Lambda$ decays in both 
$q\to 0$ and $q\to\infty$ sufficiently fast. 

Now, Eq.\ (\ref{eq:A}) can be rewritten in terms of $S_\Lambda$ in 
Eq.\ (\ref{eq:B}), (See ref.\cite{rf:MOR1})
\be
\exp\left(W[J]\right)=
\int D\phi_<\exp\left(-\Delta S_0^\Lambda
-S_\Lambda[\phi,J]\right).
\label{eq:C}
\ee
Putting $J=0$, Eq.\ (\ref{eq:C}) is nothing but the definition 
of the Wilsonian effective action $\Delta S_0^\Lambda+S_\Lambda[\phi,0]$. 
Evidently, if we put $\phi_<=0$ in Eq.\ (\ref{eq:B}) then $S_\Lambda$ 
becomes the generator of the connected Green's functions with 
the infra-red cutoff $W_\Lambda[J]=-S_\Lambda[0,J]$ \cite{rf:MOR1}. 
By shifting $\phi_>\to\phi_>-\phi_<$ and setting $J=0$ in 
Eq.\ (\ref{eq:B}), we also find \cite{rf:MOR1},
\be
W_\Lambda[P_\Lambda^{-1}\cdot\phi_<]=
\F{1}{2}\phi_<\cdot P_\Lambda^{-1}\cdot\phi_<
-S_\Lambda[\phi_<,0].
\label{eq:D}
\ee
Hereafter, we write $S_\Lambda[\phi]=S_\Lambda[\phi,0]$. 

The above cutoff $\theta$ is called the `multiplicative cutoff', because 
we multiplied it to the kinetic term of $\phi$. 
$\Delta S$ is called `additive', since the inverse cutoff propagator 
is given by $C^{-1}(q^2)$ in Eq.\ (\ref{eq:cutofffun}) plus the 
ordinary kinetic term $q^2$. $\theta(x)\equiv\theta_\Lambda^{\Lambda_0}(x)$ 
above is written in terms of $C(x)$ as $\theta(x)=xC(x)/(1+xC(x))$. 
The relation between these two cutoff schemes: the multiplicative 
cutoff and the additive cutoff is given as follows. 
The bare actions of both schemes obviously satisfy 
$S_{\Lambda_0}^{\mbox{\tiny add}}=\F{1}{2}\int d^dx(\D\phi)^2
+S_{\Lambda_0}^{\mbox{\tiny multi}}$, where 
$S_{\Lambda_0}^{\mbox{\tiny add}}$ and 
$S_{\Lambda_0}^{\mbox{\tiny multi}}$ are the bare actions with 
the additive cutoff and with the multiplicative cutoff respectively. 
By the definition, 
the generator of the connected Green's functions $W_\Lambda[J]$ is 
common for the both schemes. Thus, it immediately follows,
\be
S_\Lambda^{\mbox{\tiny multi}}[P_\Lambda\cdot J]
-\F{1}{2}J\cdot P_\Lambda\cdot J
=S_\Lambda^{\mbox{\tiny add}}[C\cdot J]
-\F{1}{2}J\cdot C\cdot J.
\label{eq:E}
\ee
We will employ the multiplicative cutoff scheme, 
since it is convenient to investigate the sharp cutoff limit of the 
RG flow equations. 

\subsection{Flow equations}

\begin{figure}[htb]
\epsfxsize=0.7\textwidth
\centerline{\epsfbox{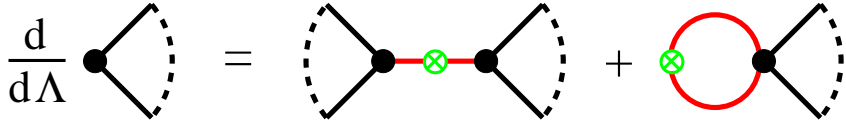}}
\caption{%
The diagrams of the Polchinski equation (\ref{eq:G}). 
The crossed circles and the filled cricles correspond to 
$\D P_\Lambda/\D\Lambda$ and the vertices of $S_\Lambda$ respectively. 
}
\label{fig:P}
\end{figure}

Setting $\phi_<=0$ and differentiating the boths sides of Eq.\ (\ref{eq:B}) 
with respect to $\Lambda$, we get the RG flow equation for 
$W_\Lambda[J]$,
\be
\F{\D}{\D\Lambda}W_\Lambda=-
\F{1}{2}W_\Lambda'\cdot \F{\D}{\D\Lambda}P_\Lambda^{-1}
 \cdot W_\Lambda'-
\F{1}{2}{\bf tr}\F{\D}{\D\Lambda}P_\Lambda^{-1}
 \cdot W_\Lambda'',
\label{eq:F}
\ee
where the prime denotes the derivative with respect to the source $J$. 
By using Eq.\ (\ref{eq:D}), we also get,
\be
\F{\D}{\D\Lambda}S_\Lambda=-
\F{1}{2}S_\Lambda'\cdot \F{\D}{\D\Lambda}P_\Lambda
 \cdot S_\Lambda'+
\F{1}{2}{\bf tr}\F{\D}{\D\Lambda}P_\Lambda
 \cdot S_\Lambda''.
\label{eq:G}
\ee
This is the famous `Polchinski equation' \cite{rf:RGE}. 
This equation may be represented diagramatically as in Fig.\ \ref{fig:P}.

Next, we scale all the dimensionful quantities in terms of the infra-red 
cutoff $\Lambda$, i.e. $\phi=\Lambda^{d_\phi}\hat\phi$, 
$\Bp=\Lambda\hat\Bp$ and ${\cal L}_\Lambda=\Lambda^d\hat{\cal L}_t$,
where $d_\phi=(d-2)/2$ is the canonical dimension of the field,
$t=\ln\Lambda_0/\Lambda$ is the cutoff scale factor
and $\hat{\cal L}_t$ is the Lagrangian density. We also 
write the dimensionless Wilsonian effective action as
$\hat S_t[\hat\phi]=\int d^d\hat x\hat{\cal L}_t(\hat\phi)$. Then, 
we get
\be
\Lambda\F{\D}{\D\Lambda}S_\Lambda=
-\Lambda^d\left(\F{\D}{\D t}+d_\phi\Delta_\phi+\Delta_\D-d\right)
\hat S_t,
\label{eq:H}
\ee
where $\Delta_\phi$ and $\Delta_\D$ count the degree of the fields and 
that of the derivatives $\D_\mu$ respectively. 
The initial boundary condition of Eq.\ (\ref{eq:G}) is given by 
the bare action, $S_{\Lambda_0}$. 

We can derive the RG flow equation for the Legendre effective action 
with the infra-red cutoff $\Gamma_\Lambda[\phi]$ given by the Legendre 
transform of $W_\Lambda[J]$. 
\be
W_\Lambda[J]=J\cdot\phi-\Gamma_\Lambda[\phi]
+\F{1}{2}\phi\cdot\left(P_\Lambda^{-1}-P_{\Lambda=0}^{-1}
\right)\cdot\phi,
\label{eq:I}
\ee
where $\phi$ is given by $\phi=\delta W_\Lambda/\delta J$. 
After the Legendre transformation, 
the RG flow equation for $\Gamma_\Lambda[\phi]$ can be read,
\be
\F{\D}{\D\Lambda}\Gamma_\Lambda[\phi]=
\F{1}{2}{\bf tr}\F{\D}{\D\Lambda}P_\Lambda^{-1}
 \cdot \left(P_\Lambda^{-1}-P_{\Lambda=0}^{-1}+
\Gamma''\right)^{-1}.
\label{eq:J}
\ee

\begin{figure}[htb]
\epsfxsize=0.7\textwidth
\centerline{\epsfbox{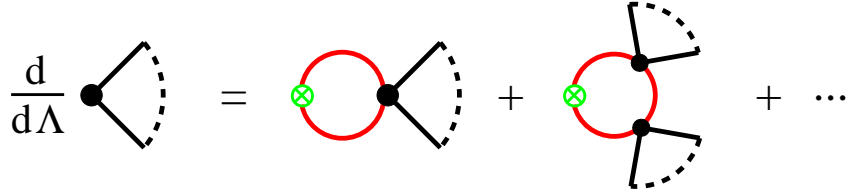}}
\caption{%
The diagrams of Eq.\ (\ref{eq:J}). The filled circles here 
correspond to the vertices of $\Gamma_\Lambda$. 
The dots denote the higher terms in the vertex of $\Gamma_\Lambda$. 
}
\label{fig:L}
\end{figure}

The initial condition of Eq.\ (\ref{eq:J}) is given by 
$\Gamma_{\Lambda_0}=S_{\Lambda_0}$, because all quantum corrections 
varnish at $\Lambda=\Lambda_0$. Since $\Gamma_\Lambda[\phi]$ 
is composed of the one particle irreducible diagrams, the 
diagrams corresponding to Eq.\ (\ref{eq:J}) contain no tree diagrams, 
as is shown in Fig.\ \ref{fig:L}. 
By the definition, $\Gamma_\Lambda[\phi]$ coincides with the ordinary 
Legendre effective action at $\Lambda=0$, i.e. 
$\Gamma_{\Lambda=0}[\phi]=\Gamma[\phi]$. One can find the RG flow 
equation for the dimensionless effective action $\hat\Gamma_t
[\hat\phi]$ by the same manner as we did for the Polchinski one. 

\subsection{Wegner-Houghton equation}

We start from the following partition function,
\be
Z=\int_{p\le\Lambda} D\phi
\exp\left(-\F{1}{2}\phi\cdot P^{-1}\cdot\phi-S_{\Lambda}[\phi]
\right),
\label{eq:K}
\ee
where the support of $\phi$ is restricted to $p\le\Lambda$. We shift 
the quadratic part, the $P^{-1}$ term, from the Wilsonian effective 
action, since it is subtracted also in $S_{\Lambda}$ given by Eq.\ (\ref{eq:B}).

Let us integrate 
out the modes with momenta $\Lambda-\delta\Lambda<p\le\Lambda$, 
we call these modes the `shell modes' and write as $\phi_{\mbox{\small s}}$. 
Expanding the action $S_\Lambda[\phi+\phi_{\mbox{\small s}}]$ in 
the shell modes $\phi_{\mbox{\small s}}$, we have
\be
S_\Lambda[\phi+\phi_{\mbox{\small s}}]=
S_\Lambda[\phi]+\phi_{\mbox{\small s}}\cdot S_\Lambda^{(1)}[\phi]
+\F{1}{2}\phi_{\mbox{\small s}}^2\cdot S_\Lambda^{(2)}[\phi]+\cdots,
\label{eq:L}
\ee
where the superscript $(n)$ 
denotes the $n$-th functional derivative with respect to 
the shell mode. We can regard the Taylor coefficients 
$S_\Lambda^{(n)}[\phi]$ to the field ($\phi$) dependent vertices.
The quantum fluctuations of the shell modes can be incorporated by 
perturbative expansion. 

\begin{figure}[htb]
\epsfxsize=0.7\textwidth
\centerline{\epsfbox{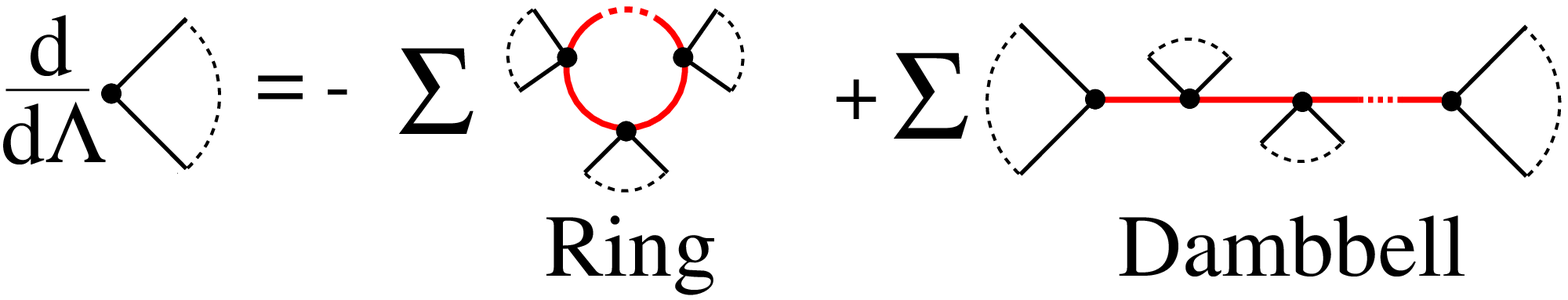}}
\caption{%
The diagrams of Eq.\ (\ref{eq:N}). The filled circles 
correspond to the vertices of $S_\Lambda$. 
}
\label{fig:WH}
\end{figure}

For infinitesimal $\delta\Lambda$, 
the leading corrections, i.e. of order $\delta\Lambda$, 
come from less than or equal to one loop diagrams. The higher 
($n\ge 2$) loops diagrams do not contribute to the leading order in 
$\delta\Lambda$, since every loop integral introduces the factor 
$\delta\Lambda$. Furthermore, each articulation line also brings 
the factor $\delta\Lambda$. Hence, the leading corrections are found 
to be the Feynman diagrams with only one propagator which is 
either an articulation line or a loop one. Taking into account this 
constraint, the higher vertices $S_\Lambda^{(n\ge 3)}[\phi]$ cannot appear 
and can be dropped in Eq.\ (\ref{eq:L}). 
After performing the Gaussian integration of the shell modes, we get a
coarse-grained action 
$S_{\Lambda-\delta\Lambda}[\phi]$ of the low energy modes 
$\phi(\Bp)$ : $p\le\Lambda-\delta\Lambda$,
\be 
S_{\Lambda-\delta\Lambda}[\phi]=S_{\Lambda}[\phi]
-\F{1}{2}\delta\Lambda S_\Lambda'\cdot
\left(P^{-1}+S_\Lambda''\right)^{-1}\cdot S_\Lambda'
+\F{1}{2}\delta\Lambda {\bf tr}\ln
\left(P^{-1}+S_\Lambda''\right),
\label{eq:M}
\ee
where prime denotes the functional derivative with respect to 
$\phi_{\mbox{\small s}}$. 
By letting $\delta\Lambda\to 0$, we find 
\be
\F{\D}{\D\Lambda}S_\Lambda=\F{1}{2}S_\Lambda'\cdot
\left(P^{-1}+S_\Lambda''\right)^{-1}\cdot S_\Lambda'
-\F{1}{2}{\bf tr}\ln
\left(P^{-1}+S_\Lambda''\right).
\label{eq:N}
\ee
This is called the `Wegner-Houghton equation' \cite{rf:RGE}. 
The diagrams corresponding to Eq.\ (\ref{eq:N}) are shown in 
Fig.\ \ref{fig:WH}.

\section{General Aspects of the Cutoff Scheme Dependence}
\label{sec:3}

In this section, we discuss some general aspects on the cutoff scheme
dependence of the ERG. In the first two subsections we show that the formal
relation among the Wilsonian effective actions with different 
cutoff schemes can be egarded as the coordinate transformation 
on the theory space. Therefore it immediately follows that 
the critical exponents, which are the macroscopic physical quantities of
the phase transition, do not suffer from the cutoff scheme dependence.
In the remaining subsections the scheme dependence of the renormalized
trajectories in the infra-red asymptotic region is examined. 
It is shown that the scheme dependence disappears form the cutoff 
Legendre effective action in this region, while not from the Wilsonian effectiveaction.

\subsection{Coordinate Transformation on the Theory Space}

The Polchinski RG equation\cite{rf:RGE} for a single scalar theory can be 
rewritten,
\begin{eqnarray}
&&\left( \F{\D}{\D t}+d_\phi \phi\cdot\F{\delta}{\delta\phi}
+\int\F{d^d q}{{\left( 2 \pi\right)}^d}\phi\left( \Bq\right)
q_\mu\F{\D }{\D q_\mu}\FDM{\Bq}
\right) \exp{\left( -S_t\left[ \phi \right]
\right)}\nonumber\\
&&\qquad\qquad\qquad\;=
\int\F{d^d q}{{\left( 2 \pi\right)}^d}\FDM{\Bq}
\left( \F{\D}{\D q^2} \theta\left( q^2 \right)\right)
\FDM{-\Bq}\exp{\left( -S_t\left[ \phi \right]\right)},
\label{eq:polchi}
\end{eqnarray}
where, for the convenience, we write the
Fourier transform of the functional derivative with respect to
$\phi(x)$ by
\be
\FDM{\Bq}\equiv\int d^dx e^{i\Bq\cdot\Bx}\FD{\Bx}=
(2\pi)^d\FD{\Bq}.
\ee
In Eq.\ (\ref{eq:polchi}), $\theta\left(q^2 \right)$ 
denotes the cutoff function,
which is given in Fig.\ \ref{fig:1} for example. 
Note that, the momentum derivative operating to the effective 
action $S_t\left[\phi\right]$ in the first line of Eq.\ (\ref{eq:polchi}) 
does not operate to the delta function 
$\delta(\Sigma\bf{q_i})$ representing momentum conservation.
Hence it operates to the effective action as
\be
\int\F{d^d q}{{\left( 2 \pi\right)}^d}\phi\left( \Bq\right)
q_\mu\F{\D }{\D q_\mu}\FDM{\Bq}
S_t\left[ \phi \right] =
\left( \Delta_\D -d \right) S_t\left[ \phi \right],
\nonumber
\ee
where $\Delta_\D$ counts the degree of derivatives.

Let us consider the coordinate transformation of the theory space:
$S_t\left[ \phi \right] \to \widetilde S_t\left[ \phi \right]$,
given by the following transformation:
\begin{equation}
\exp{\left( -\widetilde S_t\left[ \phi \right] \right)}=
\exp{\left( \F{1}{2}\delta D \right)}
\exp{\left( -S_t\left[ \phi \right] \right)},
\label{eq:trans}
\end{equation}
where $\delta D$ is given by
\begin{equation}
\delta D =
\int\F{d^d q}{{\left( 2 \pi\right)}^d}
\FDM{\Bq}
\cdot\F{1}{q^2}\delta \theta\left( q^2 \right)\cdot
\FDM{-\Bq}.
\end{equation}
Since $\delta D$ is independent of the cutoff scale $t$, 
this transformation is in fact a coordinate transformation on the 
theory space\footnote{In this paper, 
we naively assume that the coordinate transformation given by 
Eq.\ (\ref{eq:trans}) is well-defined. Since both infra-red and ultra-violet 
regions are regularized, the perturbative expansion of Eq.\ (\ref{eq:trans}) is 
finite in all orders.}. 
Therefore the critical exponents obtained by the RG technique 
are invariant under the transformation (\ref{eq:trans}). 
(See Ref.\ \cite{rf:WEIN}.) 

Indeed, by this transformation the cutoff function $\theta(q^2)$ is changed 
to $\theta(q^2)+\delta \theta(q^2)$ in the Polchinski RG equation. 
\begin{figure}[htb]
\epsfxsize=0.7\textwidth
\centerline{\epsfbox{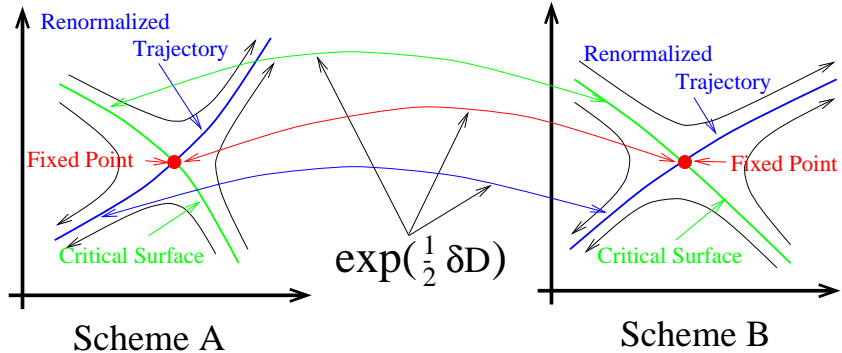}}
\caption{%
The coordinate transformation (\ref{eq:trans}) changes 
the cutoff scheme (A) to another cutoff scheme (B) in the Polchinski equation. 
}
\label{fig:2}
\end{figure}
Operating $\exp({\delta D/2})$ to both sides of Eq.\ (\ref{eq:polchi})
and using the commutation relation,
\be
\left[d_\phi \phi\cdot\F{\delta}{\delta\phi}
+\int\F{d^d q}{(2\pi)^d}\phi\left(\Bq\right)q^\mu
\F{\D}{\D q^\mu}\FDM{\Bq},\F{\delta D}{2}\right]
=\int\F{d^d q}{(2 \pi)^d}
\FDM{\Bq}
\left( \F{\D\delta \theta\left( q^2 \right)}{\D q^2}\right)
\FDM{-\Bq},
\label{eq:commute}
\ee
one can realize that the effective action $\widetilde S_t[\phi]$ 
just satisfies the Polchinski RGE with the cutoff scheme 
$\theta(q^2)+\delta \theta(q^2)$. 
Therefore, if $S_t[\phi]$ is a solution of Eq.\ (\ref{eq:polchi}) 
with cutoff scheme $\theta(q^2)$, then $\widetilde S_t[\phi]$, defined
by Eq.\ (\ref{eq:trans}), is also a solution for another cutoff scheme 
$\theta(q^2)+\delta \theta(q^2)$. 
The transformation (\ref{eq:trans}) maps the fixed point, the critical surface 
and the renormalized trajectories to those given in another scheme. 
(See Fig.\ \ref{fig:2}.) 
However the critical exponents at the fixed points are scheme independent.

\subsection{Wave-function Renormalization and Additive cutoff}

In order to extract the anomalous dimension it is more convenient to
employ the additive cutoff instead of the multiplicative one for 
following two reasons. 
1). In the multiplicative case the part of the kinetic term of 
the Wilsonian effective action is stolen by the inverse cutoff 
propagator $q^2\left(\theta(q^2)\right)^{-1}$. 
2). We can elminate the wave-function renormalization factor $Z_\phi$ 
in RG equation by rescaling the cuotoff function 
$C\left(q^2\right)$ to $Z_\phi^{-1}C\left(q^2\right)$
in the additive case, and the anomalous dimension $(\eta)$
of the field may be explicitly extracted. 

The additive cutoff is introduced by Eq.\ (\ref{eq:cutofffun}) and 
the Wilsonian effective actions with two cutoff schemes are related 
by the relation (\ref{eq:E}), i.e.
\be
S^{\mbox{\tiny multi}}[\phi]=
S^{\mbox{\tiny add}}[(1-\D^2C)\phi]
+\F{1}{2}\int d^dx\phi\D^2(1-\D^2C)\phi,
\ee
where $S^{\mbox{\tiny multi}}[\phi]$ is the effective action with
the multiplicative cutoff. 
In the multiplicative cutoff case, we drop the part of the kinetic
term of (the interaction part of) the effective action. 
It is rather convenient to include the kinetic term in the effective action 
completely in extracting the anomalous dimension. 
The additive cutoff $C(x)$ in Eq.\ (\ref{eq:cutofffun}) and 
the multiplicative cutoff $\theta(x)$ are related by 
$\theta(x)=xC(x)/(1+xC(x))$. 

Let us rescale the field $\phi$ to $\hat\phi=Z_\phi^{\F{1}{2}}\phi$, 
where $Z_\phi$ is the wave-function renormalization factor. 
If we also rescale the cutoff function as,
\begin{equation}
C^{-1}\left( q^2/\Lambda^2 \right)\longrightarrow
Z_\phi C^{-1}\left( q^2/\Lambda^2 \right),
\label{eq:newcut}
\end{equation}
then the explicit $Z_\phi$ dependence of the RG flow equation can be 
eliminated. 
The RG flow equation depends on $Z_\phi$ only through the anomalous dimension 
$\eta$ defined by the consistency condition, 
i.e. the kinetic term should be unity at each scale. 

Consequently, $\eta$ becomes the function of the coupling constants. 
It means that the beta-function of $Z_\phi$ is given by, 
\be
\F{\D}{\D t}Z_\phi=\eta(g_i)Z_\phi,
\ee
where $\{g_i\}$ is a coordinate system on the theory space. 
In this coordinate system, the RG 
beta-functions have the following structure:
\be
\Omega_{ij}(g)=\F{\D\beta_i}{\D g_j}(g)=0\quad
{\rm for}\quad g_j=Z_\phi,
\ee
because the beta-functions have no $Z_\phi$ dependence except for $\beta_Z$. 
Such a parametrization is called the `perfect coordinate' in Ref.\ \cite{rf:SO}.
For the dimensionless Wilsonian effective action, the RG flow equation becomes,
\begin{eqnarray}
&&\left( \F{\D}{\D t}+d_\phi \phi\cdot\F{\delta}{\delta\phi}
+\int\F{d^d q}{{\left( 2 \pi\right)}^d}\phi\left( \Bq\right)
q_\mu\F{\D }{\D q_\mu}\FDM{\Bq}
\right) \exp{\left( -S_t\left[ \phi \right]
\right)}\label{eq:version2}\\
&&\qquad\qquad\qquad\quad\;=-
\int\F{d^d q}{{\left( 2 \pi\right)}^d}\FDM{\Bq}
\left( q^2\F{\D}{\D q^2} C
+\F{1}{2}(\eta-2) C\right)
\FDM{-\Bq}\exp{\left( -S_t\left[ \phi \right]\right)},
\nonumber
\end{eqnarray}
where $d_\phi$ and $\eta=\dot Z_\phi/Z_\phi$ are the physical scaling 
dimension of the field; $d_\phi =(d+\eta-2)/2$ and the anomalous dimension of 
$\phi$ respectively. One can easily extend the scheme dependence
relations given by Eq.\ (\ref{eq:commute}) etc. to this type of RGE.

However one may wonder whether the coordinate transformation induced by 
Eq.\ (\ref{eq:newcut}) is well-defined or not. 
Suppose that $C(x)$ is a polynomial i.e. $C(x)=x^k$. Since the cutoff 
$\Lambda$ appears only though the cutoff function, 
\be
{\rm e}^{-S_\Lambda[\phi]}\equiv
\int D\;\phi_>
{\exp}\;\Big \{-\F{Z_\phi}{2}\phi_>
\Lambda^2C^{-1}(-\D^2/\Lambda^2)\phi_>
-S[\phi_>+\phi]\Big \},
\ee
we can eliminate $Z_\phi$ by shifting the cutoff; 
$Z_\phi\Lambda^2C^{-1}(\F{p^2}{\Lambda^2})
={\Lambda'}^2C^{-1}(\F{p^2}{{\Lambda'}^2})$ 
where $\Lambda=Z^{1/(2k-2)}_\phi{\Lambda'}$. Therefore 
the Wilsonian effective action  $\widetilde S_\Lambda[\phi]$
with a cutoff scheme $C^{-1}$ can be written in terms of 
$S_\Lambda[\phi]$, 
\begin{eqnarray}
&&\widetilde S_{\Lambda'}[\phi]=S_\Lambda[\phi]
=S_{\Lambda'}[\phi]+
\int_{\Lambda'}^\Lambda d\bar\Lambda\F{\D}{\D\bar\Lambda}
S_{\bar\Lambda}[\phi],\nonumber\\
&&\qquad\;\;=S_{\Lambda'}[\phi]
+\delta f(Z_\phi)\Lambda'\F{\D}{\D\Lambda'}
S_{\Lambda'}[\phi]
+\F{1}{2!}(\delta f(Z_\phi)\Lambda')^2\F{\D^2}{\D{\Lambda'}^2}
S_{\Lambda'}[\phi]+\cdots,
\label{eq:expanW}
\end{eqnarray}
where $\delta f(Z_\phi)$ is given by $\Lambda-\Lambda'=
\delta f(Z_\phi)\Lambda'$. 
In our case, $\delta f(Z_\phi)=Z^{1/(2k-2)}_\phi-1$. 
The derivative with respect to $\Lambda'$ in Eq.\ (\ref{eq:expanW}) 
will be eliminated by the RG flow equation. Thus one can find 
the coordinate transformation between $\widetilde S_\Lambda[\phi]$ 
and $S_\Lambda[\phi]$. 
Since all the loop momentum integrals are regularized in the both regions of
infra-red and ultra-violet, Eq.\ (\ref{eq:expanW}) gives the well-defined
coordinate 
transformation at all orders in the Taylor expansion. 

In the more general case, we cannot eliminate $Z_\phi$ by shifting 
the cutoff. However, since the change of $\theta(x)=xC(x)/(1+xC(x))$ 
induced by the change of the cutoff function $C^{-1}(x)$ is concentrated 
in the finite region of the momentum $x=p^2$, it also gives the well-defined 
coordinate transformation equivalent to Eq.\ (\ref{eq:trans}).

Consequently, the critical exponents given by Eq.\ (\ref{eq:polchi}) and by
Eq.\ (\ref{eq:version2}) 
are completely the same, since these formulations 
can be understood as the difference of the coordinate systems on the 
theory space. 

\subsection{Asymptotic Region of the Renormalized Trajectory}

In this and the next subsections we discuss the cutoff scheme dependence
of the renormalized trajectories in the `asymptotic region' 
$\Lambda\ll M_R$, where $M_R$ is the renormalized mass of $\phi$. 
We note that Eq.\ (\ref{eq:trans}) may be rewritten as follows. 
Let $S_t^{(n)}$ be the vertices of the effective action $S_t[\phi]$, and 
 $\widetilde{W}_t$ be sum of the connected diagrams composed of 
the propagator $(\delta P^{-1}+S_t^{(2)})^{-1}$ and the vertices $S_t^{(n)}$ 
( $n>2$ ), where $\delta P(q^2)$ is the cutoff propagator 
i.e. $\delta P(q^2)\equiv\F{1}{q^2}\delta \theta(q^2)$.
Then it is easily found that (See appendix),
\begin{equation}
\exp{\left( \F{1}{2}\delta D[\delta/\delta\phi] \right)}
\exp{\left( -S_t\left[ \phi \right] \right)}=
\exp{\left(-\F{1}{2}\phi\cdot\delta P^{-1}\cdot\phi
+\widetilde{W}_t[\delta P^{-1}\cdot\phi]\right)}.
\label{eq:connect}
\end{equation}
The cutoff scheme dependence of the renormalized trajectory is given 
by Eq.\ (\ref{eq:connect}). (See Fig.\ \ref{fig:2}.) 

If the RG flows of the dimensionful quantities freeze when $t\to\infty$, 
then all dimensionless coupling 
$g_i(t)$ should behave $g_i(t)\sim g_i^R e^{d_it}$ as $t\to\infty$ 
with some finite dimensionful coupling $g_i^R$, where $d_i$ is the 
canonical dimension of $g_i$. 
Here, we take $M_R$ to be a unit of the mass scale:
$t=\ln M_R/\Lambda$. Hereafter we call such a  region `the freezing region',
if it exists. As is aeen below the asymptotic region is not always the
freezing region.

In the asymptotic region, it can be realized that the loop diagrams in 
Eq.\ (\ref{eq:connect}) 
are found to be suppressed compared with the tree diagrams. 
It is seen by the following arguments. Let us consider the 
Feynman diagram with $N_I$ internal lines, $N_E$ external legs and 
$N_V$ vertices. By comparing the canonical dimension of each operator 
in the both sides of Eq.\ (\ref{eq:connect}), we find the following factor,
\be
\exp\{\Delta t\}\equiv
\exp\left\{\left(
dN_V-d_\phi(N_E+2N_I)-N_D^{(1)}
-2N_I
-(d-d_\phi N_E)+N_D^{(2)}
\right)t\right\},
\ee
where $N_D^{(1)}$  is the total degree of the 
external momenta (or the derivatives) 
of the $N_V$ vertices in the Feynman diagram, 
and $N_D^{(2)}$ is that of the subset of $N_D^{(1)}$ derivatives 
which operate to the field $\phi(x)$. 
The first three terms of $\Delta$ come from the canonical dimensions of 
the vertices, the next one is from propagators and the last two terms are 
the dimension of $N_E$ point vertex with $N_D^{(2)}$ derivatives. 
$N_D^{(1)}$ and $N_D^{(2)}$ satisfy the relation $N_D^{(1)}\ge 
N_D^{(2)}$, since the number of derivatives only decreases by the loop 
integration
\footnote{Some of the derivatives will be replaced by the 
loop momenta $q\sim 1$ which does not contribute to $\Delta$.}. 
We do not expand the propagators with respect to the external momenta, since 
we are interested only in the scaling behavior of the Feynman diagrams 
in which each propagator gives a negative power of the external momenta 
and brings the factor $\exp(-2t)$
\footnote{Namely we do not perform 
the derivative expansion here The asymptotic behavior in the derivative 
expansion will be discussed in \S~6.
}. 
The loop integration does not change the factor $\Delta$, because the support 
of $\delta\theta(\hat p^2)$ is concentrated in the small region around 
$\hat p^2\sim 1$. The factor $\Delta$ describes 
the response of the shrinkage of the loop momentum integral region. 
The diagrams with $\Delta<0$ are dropped in Eq.\ (\ref{eq:connect}). 
{\it Nota Bene} the massive field decouples in the asymptotic region 
not due to its large (dimensionless) mass but due to the shrinkage 
of the loop momentum integral region. 

One can rewrite  $\Delta$ by using the number of the loops $L$,
\be
\Delta[L]=-dL-(N_D^{(1)}-N_D^{(2)})\le -dL,
\ee
where we used the relation,
\be
N_I-N_V+1=L.
\ee
Hence all loop corrections i.e. `quantum' corrections are relatively 
suppressed by the factor $\exp(-dLt)$ compared with the tree diagrams. 
Since the number of derivatives is decreased by only loop integration, 
$N_D^{(1)}=N_D^{(2)}$ in the tree diagrams. Therefore all the tree diagrams
survive. 
Now we can conclude that in the asymptotic region $\widetilde W_t$ 
in Eq.\ (\ref{eq:connect}) is consist of tree diagrams only. 

\begin{figure}[htb]
\epsfxsize=0.78\textwidth
\centerline{\epsfbox{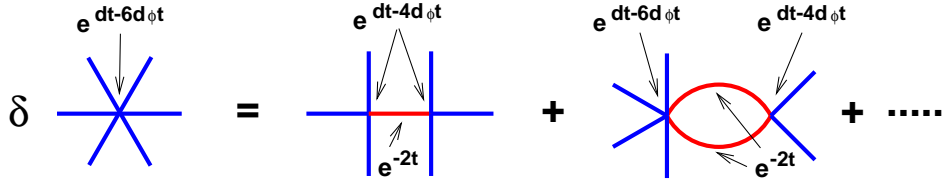}}
\caption{%
The examples of the factor $\Delta$ are shown. In this figure, 
$d_\phi$ is the canonical dimension of the field i.e. $d_\phi=(d-2)/2$. 
The dots denote the higher order corrections. 
}
\label{fig11}
\end{figure}

In Fig.\ \ref{fig11} we show examples of the Feynman 
diagrams and their suppression factors. 
In these examples, the factor $\Delta$ of first tree diagram is 
$\Delta=2(d-4d_\phi)-2-(d-6d_\phi)=0$ and that of the second loop diagram is
$\Delta=2d-(4+6)d_\phi-4-(d-6d_\phi)=-d<0$. The later diagram is 
suppressed in comparison with the six points vertex itself. 

The above observation holds also for the RG flows, 
in which $\delta P$ is induced by lowering the cutoff $\Lambda$. 
Hence the RG flows of the dimensionful couplings of the 1PI building 
blocks of the Wilsonian effective action freeze in the asymptotic region, 
however the Wilsonian effective action itself does not. 
The later conflicts to our assumption made before, i.e. $g_i(t)\sim g_i^R
e^{d_it}$ with fixed $g_i^R$. Thus {\it there is no freezing region on the RG 
flow diagram of the Polchinski equation}. 

The coordinate transformation 
$\Delta_{\delta P}[S_t]$ in the asymptotic region is written by,
\be
\Delta_{\delta P}[S_t[\phi]]=\F{1}{2}\phi\cdot\delta P^{-1}
\cdot\phi-\widetilde W_{\rm tree}[\delta P^{-1}\phi],
\ee
where $\widetilde W_{\rm tree}$ is the tree part of the connected diagrams. 
It can be easily realized that $\widetilde W_{\rm tree}$ is given 
by the Legendre transform of the effective action $S_t$, since the 1PI 
part of $\widetilde W_{\rm tree}$ is nothing but the `Legendre 
effective action' $\widetilde\Gamma_t$.
($\widetilde\Gamma_t$ should not be confused with 
$\Gamma_\Lambda$ given in Eq.\ (\ref{eq:I}), 
which is equal to the effective action $S_t$ in the asymptotic region
\footnote{All the loop (quantum) corrections for the 1PI vertices 
are dropped in the asymptotic region. It means 
$\widetilde\Gamma_t=S_t$.}. )
Therefore we find,
\be
\widetilde W_{\rm tree}[J]=\bar\phi\cdot J
-S_t[\bar\phi]-\F{1}{2}\bar\phi\cdot\delta P^{-1}
\cdot\bar\phi.
\label{eq:eq11}
\ee
We add the last term of r.h.s. of Eq.\ (\ref{eq:eq11}) to the effective action, 
because $\widetilde W_t$ in Eq.\ (\ref{eq:connect}) consists of the connected 
diagrams with bare action 
$S_t[\phi]+\F{1}{2}\phi\cdot\delta P^{-1}\cdot\phi$. 
Here $J$ and $\bar\phi$ satisfy the relations,
\be
J-\delta P^{-1}\bar\phi=\F{\delta}{\delta\bar\phi}S_t,\qquad
\bar\phi=\F{\delta}{\delta J}\widetilde W_{\rm tree}.
\ee
One can find the coordinate transformation $\Delta_{\delta P}$ as
\be
\Delta_{\delta P}[S_t[\phi]]=
\F{1}{2}S'_t[\bar\phi]\cdot\delta P\cdot
S'_t[\bar\phi]+S_t[\bar\phi],\label{eq:delta1}
\ee
where the prime denotes the functional derivative with respect to 
$\bar\phi\left(\equiv\phi-\delta P S'_t[\bar\phi]\right)$. 
One can also easily check the following relations,
\begin{eqnarray}
&&\Delta_{\delta P_1}[\Delta_{\delta P_2}[S_t]]
=\Delta_{\delta P_1+\delta P_2}[S_t],\\
&&\Delta_{\delta P}[\Delta_{-\delta P}[S_t]]=S_t.
\end{eqnarray}

The scheme dependence of the Wilsonian effective action may be understood 
as follows. As discussed in Ref.\ \cite{rf:MOR1}, 
the Wilsonian effective action is consist of two different elements. 
In the high energy region, the vertices of the effective action give the
connected Green's functions. 
In the low energy region, they coincide with those of the 1PI effective 
action, since all the articulation lines carry the infra-red 
cutoff.\cite{rf:MOR1} 
In the boundary of these regions, two quantities are connected to each 
other by the cutoff function. Therefore, the Wilsonian effective action 
is scheme dependent even after removing the cutoff. 
This scheme dependence turns out to be an obstacle in the approximated 
analyses. (See Sec.\ \ref{sec:6}.)

\subsection{Scheme independence of the cutoff Legendre effective action}

\begin{figure}[htb]
\epsfxsize=0.7\textwidth
\centerline{\epsfbox{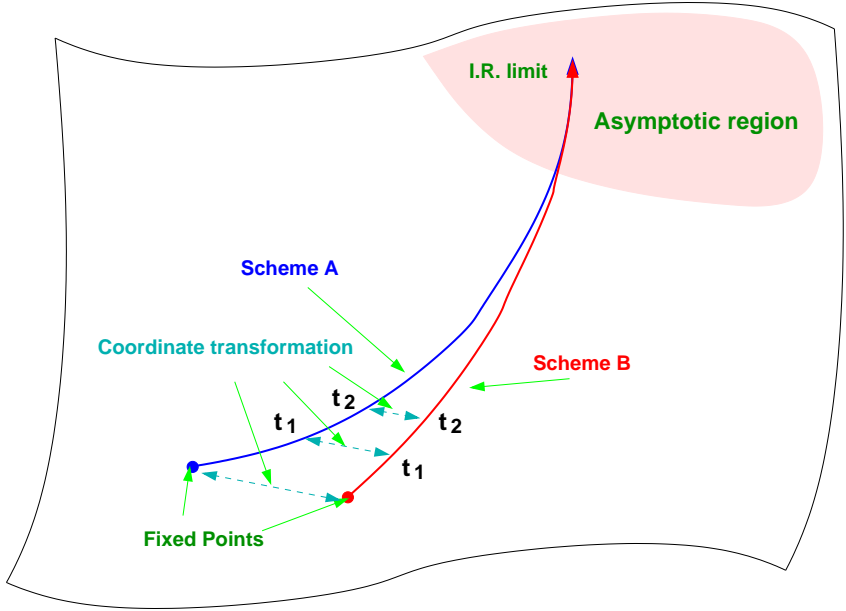}}
\caption{%
The scheme dependence of the renormalized trajectories of the 1PI vertices
(the solid lines). 
The dotted arrows are the coordinate transformation (\ref{eq:trans})
and the shadow region is the asymptotic (freezing) region. 
The dimensionful cutoff Legendre effective action is frozen in this 
region and the renormalized trajectories with the different cutoff schemes 
come close to each other.
}
\label{fig:RGflow}
\end{figure}

For the infinitesimal transformation $\delta P\ll 1$, 
the scheme dependence of the generator of the connected Green's functions 
$W_t[J]$ becomes simpler. Now $W_t[J]$ is written 
in terms of the Wilsonian effective action $S_t[\phi]$ by
\be
W_t[J]=\F{1}{2}J\cdot P_{\Lambda(t)}\cdot J-
S_t[P_{\Lambda(t)}J].
\label{eq:gfgf}
\ee
By using Eq.\ (\ref{eq:delta1}) we find 
\be
\delta W_t=\F{1}{2}W'_t\cdot
\F{\delta P_{\Lambda(t)}}{P_{\Lambda(t)}^2}
\cdot W'_t.\label{eq:delW}
\ee
Since all the loop corrections are suppressed, the 1PI parts of the 
cutoff connected 
Green's functions in the asymptotic region are completely scheme independent. 
Indeed, by the Legendre transformation
\be
W_t[J]=J\cdot\phi-\Gamma_t[\phi]-\F{1}{2}
\phi\cdot\left(P_{\Lambda(t)}^{-1}-P_{\Lambda(t)=0}^{-1}\right)\cdot\phi,
\ee
and Eq.\ (\ref{eq:delW}), one can find $\delta\Gamma_t[\phi]=0$. 
Namely, the renormalized trajectories of the 1PI vertices defined in the 
different cutoff schemes approaches to each other in the asymptotic
region, as is schematically shown in  Fig.\ \ref{fig:RGflow}. 

The coordinate transformation on the functinal space of the 1PI 
effective action $\Gamma_t[\phi]$ induced by Eq.\ (\ref{eq:trans}) 
maps the fixed points, the renormalized trajectories and the critical 
surfaces to those of the another scheme. Here we write this coordinate
transformation as $\Delta_{(A\to B)}$. 
Once the effective action of the scheme A ($\Gamma_t^A[\phi]$) 
and that of the scheme B ($\Gamma_t^B[\phi]$) satisfy the relation,
\be
\Gamma_{t_1}^B[\phi]=\Delta_{(A\to B)}[\Gamma_{t_1}^A],
\label{eq:transL}
\ee
at a certain scale $t_1$, then it holds also for the each scale $t_2$. 
(See Fig.\ \ref{fig:RGflow}.) Solving the RG flow equation, the
cutoff Legendre effective actions finally arrive at the asymptotic 
region with maintaining the same relation~(\ref{eq:transL}). 
As discussed in the last subsection $\Delta_{(A\to B)}$ reduces to
the identical mapping in the asymptotic region. Note that the RG 
flow of the dimensionful cutoff Legendre effective action is frozen 
in the asymptotic region. 

The continuum limit of the field theories are found by tuning the 
initial boundary condition of the RG equation close to the critical
surface or the fixed point and are discribed by the renormalized 
trajectories. As is seen above, the renormalized trajectories are
cutoff scheme independent in the freezing region. 
This converging property of the RG flows of the cutoff Legendre 
effective action ensures that the solutions of the RG flow equations 
become cutoff scheme independent in the continuum limit.
Needless to say, each theory must be specified by imposing the 
renormalization conditions for the renormalized couplings. 
Then other couplings are determined scheme independently.
This structure should be compaired with the scheme dependence of
the Wilsonian effective action (or the Polchinski RGE). 
It is an advantageous feature of the Legendre flow equations that
the physically meaningful results can be obtained directly. 

\section{Sharp cutoff limit of Polchinski equation}
\label{sec:4}

In this and next section, we confirm the equivalence between 
the sharp cutoff limit of the Polchinski equation and the
Wegner-Houghton equation. 
It seems that the sharp cutoff RG equation, the Wegner-Houghton RG 
\cite{rf:RGE} 
is quite different from the smooth cutoff one; the Polchinski RG. 
At first, we clarify the sharp cutoff limit of the Polchinski RGE
in this section, and confirm the equivalence to the Wegner-Houghton
equation in next section. 

Since we would like to consider the sharp cutoff limit, it is more 
convenient to write the cutoff propagator $P_\Lambda$ in terms of 
the cutoff function $\theta_\varepsilon(q,\Lambda)$. Here
$\theta_\varepsilon$ is a smooth function with respect to the momentum 
$q$, the cutoff $\Lambda$ and a smoothness parameter $\varepsilon$. 
In the limit of $\varepsilon\to 0$, $\theta_\varepsilon(q,\Lambda)$ 
becomes a step function $\theta(q-\Lambda)$, therefore, 
\be
P_\Lambda(q)=\F{1}{q^2}\theta_\varepsilon(q,\Lambda)
\stackrel{\varepsilon\to 0}{\longrightarrow}
\F{1}{q^2}\theta(q-\Lambda).
\ee
If we also introduce $\delta_\varepsilon(q,\Lambda)$ denoting 
the derivative of $\theta_\varepsilon$ with respect to the 
cutoff $\Lambda$, which satisfies
\be
-\F{\D}{\D\Lambda}\theta_\varepsilon(q,\Lambda)
=\delta_\varepsilon(q,\Lambda)
\stackrel{\varepsilon\to 0}{\longrightarrow}
\delta(q-\Lambda).\label{eq:limit}
\ee

It is pointed out in Ref.\ \cite{rf:MOR1} that one has 
to be careful for the behavior of $\theta_\varepsilon$ and $\delta_
\varepsilon$ in the sharp cutoff limit. The non-trivial and universal 
behavior of $\theta_\varepsilon$ and $\delta_\varepsilon$ is 
\be
\delta_\varepsilon(q,\Lambda)f(\theta_\varepsilon
(q,\Lambda),q,\Lambda)\stackrel{\epsilon\to0}{\longrightarrow}
\delta(\Lambda-q)\int_0^1\!\!dt\,f(t,q,\Lambda).
\label{eq:limit1}
\ee
For the derivation of the above formula, see Ref.\ \cite{rf:MOR1}. 
Note that, $\theta_\varepsilon$ with another momentum $q'\ne q$ does 
not behave as Eq.\ (\ref{eq:limit1}). 

Let $S_\Lambda^{\varepsilon=0}$ be the effective action 
for the sharp cutoff case and 
$S_\Lambda^\varepsilon$ be one for the smooth cutoff case 
with $\theta_\varepsilon$. 
These two effective actions should be related to each other 
by the formula (\ref{eq:trans}), 
\be
\exp{\left( -S_\Lambda^\varepsilon\left[ \phi \right] \right)}=
\exp{\left( \F{1}{2}\delta D \right)}
\exp{\left( -S_\Lambda^{\varepsilon=0}\left[ \phi \right] \right)},
\label{eq:scheme2}
\ee
where
\be
\delta D =
\int\F{d^d q}{{\left( 2 \pi\right)}^d}
\FDM{\Bq}
\cdot\F{1}{q^2}\left(\theta_\varepsilon(q^2)-\theta(q-\Lambda)
\right)\cdot
\FDM{-\Bq}.
\ee
Here, $S_\Lambda^\varepsilon[\phi]$ has the dependence  of
$\theta_\varepsilon$ so that we have to take into account of 
Eq.\ (\ref{eq:limit}). 
For the sake of simplicity, we write $\theta_\varepsilon(q_i^2)-
\theta(q_i-\Lambda)\equiv \Delta_i$. 

\begin{figure}[htb]
\epsfxsize=0.5\textwidth
\centerline{\epsfbox{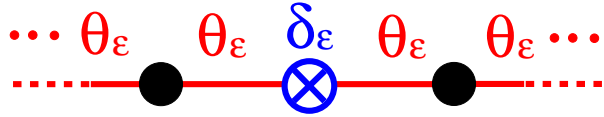}}
\caption{%
This kind of the diagrams are crucial for the sharp cutoff limit of the 
Polchinski equation. $\delta_\varepsilon$ corresponds to $\D\theta(q^2)
/\D q^2$ in the l.h.s. of Eq.\ (\ref{eq:polchi}). 
}
\label{fig:5}
\end{figure}

The cutoff functions $\theta_\varepsilon$ contributing to the non-trivial 
limit (\ref{eq:limit1}) should have common argument $q$ with that of 
$\delta_\varepsilon$ in the RG flow equation. They lie only on 
the external legs. Diagrammatically, they can be found 
in Fig.\ \ref{fig:5}. 
In this diagram, all the momentum of 
$\theta_\varepsilon(q)$ are the same as that of $\delta_\varepsilon$. 
Here, the filled circles in Fig.\ \ref{fig:5} correspond to the 
self energy $\Sigma^\varepsilon_\Lambda(q)$. They are summed up 
and construct the `Full propagator' times the inverse cutoff 
propagator i.e. 
$(q^2/\Delta)/(q^2/\Delta+\Sigma^\varepsilon_\Lambda(q))$. 
The pre-factor $q^2/\Delta$, the inverse cutoff propagator, 
is required, since the argument of $\widetilde W_\Lambda$ in 
Eq.\ (\ref{eq:connect}) is $\delta P^{-1}\phi$. 

\begin{figure}[htb]
\epsfxsize=0.4\textwidth
\centerline{\epsfbox{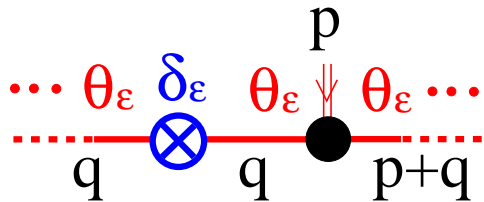}}
\caption{%
The beta-function and also the effective action have the discontinuous 
momentum dependence coming from these diagrams. 
}
\label{fig:6}
\end{figure}

In general, one can imagine other diagrams like Fig.\ \ref{fig:6} which 
the momentum $p$ flows in. In this figure, the filled circle denotes 
the multi-point 1PI vertex. 
Since the cutoff function $\theta_\varepsilon(\Bp+\Bq)$ behaves 
differently at $p=0$ in the sharp cutoff limit, 
the discontinuous momentum dependence should appear in the beta-functional. 
If the field $\phi(\Bp)$ is a smooth function, there are 
no finite contributions since the measure of the point $p=0$ is zero. 
However since $\phi(\Bp)$ may have the distribution like a VEV 
$\varphi(2\pi)^d\delta^d(\Bp)$, we separate such distributions explicitly. 
Furthermore, 
since we would like to claim the equivalence between the Wegner-Houghton 
equation and  the Polchinski equation in the sharp cutoff limit including 
these discontinuous momentum dependence, we also introduce 
the singularities like $\delta^d\left(\Bq-\Bq_i\right)$,
\be
\phi\left(\Bq\right)\longrightarrow
(2\pi)^d\delta^d\left(\Bq-\Bq_i\right)
\varphi_i+\phi\left(\Bq\right).
\label{eq:field}
\ee

Now, the effective action acquires $\varphi_i$ dependence i.e. 
$\hat{S}_\Lambda^\varepsilon=\hat{S}_\Lambda^\varepsilon[\phi,\varphi_i]$ 
and also satisfies the same formula (\ref{eq:scheme2}). 
In this case, the Feynman 
diagrams like Fig.\ \ref{fig:6} contribute to the sharp cutoff flow equation 
via the combination $\prod\varphi_i^{n_i}$ with 
$\sum n_ip_i=0,~n_i\in {\bf N}$, that is a point $p=0$ of the diagram
in Fig.\ \ref{fig:6}. The corrections from $\prod\varphi_i^{n_i}$ terms
can be absorbed by redefinition of the self energy 
$\Sigma^\varepsilon_\Lambda(q)$. Consequently we regard 
the self energy $\Sigma^\varepsilon_\Lambda$ as a $\varphi_i$ dependent
function. 

One may wonder if the  self energy $\Sigma^\varepsilon_\Lambda(q)$ 
has the discontinus 
momentum dependence, since there are $\theta_\varepsilon(q)$ 
with momenta in common with $\delta_\varepsilon$'s in the loop integrals 
giving $\Sigma^\varepsilon$. 
However, since the regions of the loop momentum integration have the vanishing
measure, the above discontinuous momentum dependence does not contribute to 
Eq.\ (\ref{eq:limit1}) at all. Similarly, all the 1PI building blocks 
smoothly approach to those for the sharp cutoff. Hence, what we must 
check is only articulation lines. Furthermore, the internal lines 
with momenta $\Bq+\Bp$ vanish, since $\Delta(q+p)$ goes to zero 
in the limit $\varepsilon\to 0$. Therefore we need to care 
the external legs only. 

\begin{figure}[htb]
\epsfxsize=0.9\textwidth
\centerline{\epsfbox{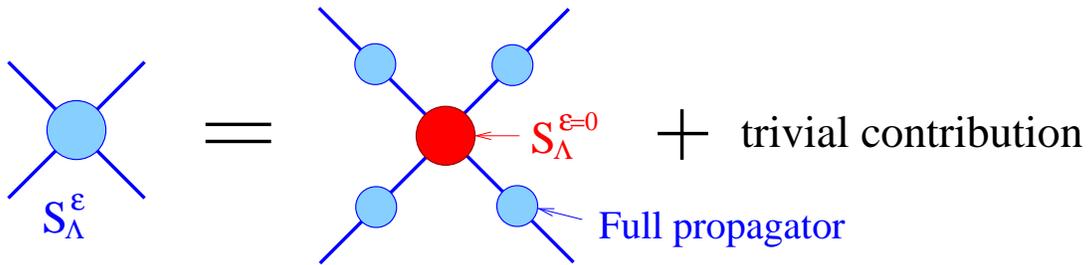}}
\caption{%
These kinds of diagrams behave as Eq.\ (\ref{eq:limit}). 
$S_\Lambda^{\varepsilon=0}$ is the vertex with the sharp cutoff
and the `full propagator' is the propagator with the smooth one. 
Other contributions shown by the `trivial contribution' 
vanish in the limit $\varepsilon\to 0$.
}
\label{fig:7}
\end{figure}

Finally, one can conclude that the relevant scheme dependence 
of $n(>2)$-point functions comes only from 
the external legs. Diagrammatically, they can be illustrated as in 
Fig.\ \ref{fig:7}. 
The scheme dependence of the two-point function $S_\Lambda^{(2)}$ is given by 
$q^2/\Delta$ minus the `Full propagator' times $(q^2/\Delta)^2$ and is
different from those of other vertices. Therefore we 
separate the two point function in the effective action and write as
\be
S_\Lambda[\phi]=\F{1}{2}\phi\cdot S_\Lambda^{(2)}\cdot\phi
+\hat S_\Lambda[\phi],
\ee
where $\hat S_\Lambda[\phi]$ is the part composed of $n(>2)$-point functions.

Taking account of Eq.\ (\ref{eq:connect}), we can extract the relevant 
scheme ($\Delta$) dependence as follows. For the two-point 
function $S_\Lambda^{(2)}$,
\be
\F{\bar\delta^2 S_\Lambda^\varepsilon}
{\bar\delta\phi(\Bp)\bar\delta\phi(\Bq)}\Big |_{\phi=0}=
\left(\F{q^2}{\Delta}\right)^2\left(\F{\Delta}{q^2}-
\F{1}{q^2/\Delta+\Sigma_\Lambda^{\varepsilon}(q)}\right)\cdot
(2\pi)^d\delta^d(\Bp+\Bq),\label{eq:two}
\ee
and for $n(>2)$-point functions $\hat S_\Lambda$,
\be
\hat{S}_\Lambda^\varepsilon[\phi]=\sum_{n\ne2}\F{1}{n!}
\prod_{i=1}^n\int\F{d^d q_i}{{\left( 2 \pi\right)}^d}
\left(\F{q_i^2/\Delta_i\cdot\phi(\Bq_i)
}{q_i^2/\Delta_i+\Sigma_\Lambda^{\varepsilon}}\right)
S_\Lambda^{\varepsilon=0}(\Bq_1,\cdots,\Bq_n)
+\cdots,
\label{eq:vertex}
\ee
where dots `$\cdots$' have no significant dependence on $\Delta$ and vanish 
in the sharp cutoff limit. As mentioned above $\Sigma_\Lambda^\varepsilon$ 
and $S_\Lambda^\varepsilon$ depend on $\varphi_i$, $e.g$. $\Sigma_\Lambda
^\varepsilon(q)=\Sigma(q,{\mbox{\small$\prod$}}\varphi_i^{n_i})$ and $\hat S_\Lambda
[\phi,{\mbox{\small$\prod$}}\varphi_i^{n_i}]$. 

Let us first see the sharp cutoff limit of Eqs.\ (\ref{eq:two}) and
(\ref{eq:vertex}). In this limit, $\Delta$ vanishes and
$\Sigma_\Lambda^\varepsilon(q)$ can be replaced 
by $\Sigma_\Lambda^{\varepsilon=0}(q)$ safely. 
For $n>2$ we easily find
\be
\hat{S}_\Lambda^{\varepsilon=0}[\phi]=\sum_{n\ne2}\F{1}{n!}
\prod_{i=1}^n\int\F{d^d q_i}{{\left( 2 \pi\right)}^d}
\phi(\Bq_i)
S_\Lambda^{\varepsilon=0}(\Bq_1,\cdots,\Bq_n).
\ee
For the two-point function, we can rewrite Eq.\ (\ref{eq:two}) as,
\be
\Sigma_\Lambda^{\varepsilon=0}(q)=q^2\Sigma_\Lambda^\varepsilon
/\left(q^2+\Delta\cdot\Sigma_\Lambda^\varepsilon\right),
\ee
where $\Sigma_\Lambda^{\varepsilon=0}$ corresponds to the two 
point function of the sharp cutoff effective action 
$S_\Lambda^{\varepsilon=0}[\phi]$,
\be
\F{\bar\delta^2 S_\Lambda^{\varepsilon=0}}
{\bar\delta\phi(\Bp)\bar\delta\phi(\Bq)}\Big |_{\phi=0}=
\Sigma_\Lambda^{\varepsilon=0}(q,{\mbox{\small$\prod$}}\varphi_i^{n_i})
(2\pi)^d\delta^d(\Bp+\Bq).
\label{eq:2pfun}
\ee

We must fix the $\theta(0)$ ambiguity before letting $\varepsilon\to 0$ 
in the RG flow equation, since $\Delta$ in Eqs.\ (\ref{eq:two}) and
(\ref{eq:vertex}) is given by 
$\Delta(q)=\theta_\varepsilon(q/\Lambda)-\theta(q-\Lambda)$ and satisfy
\be
\delta_\varepsilon(q/\Lambda)f(\Delta(q))
\stackrel{\epsilon\to0}{\longrightarrow}
\delta(q-\Lambda)\int_0^1dtf(t-\theta(0)).
\ee
We simply set $\theta(0)=0$ here. It is different from the ordinary 
convention; $\theta(0)=1/2$. This is because, we implicitly used 
$\theta(0)=0$ to derive the Wegner-Houghton equation. 
The `shell modes' $\phi_s$ integrated out by the RG transformation have 
momenta $\Lambda-\delta\Lambda<q\le\Lambda$, lower than the scale
$\Lambda$ of the effective action $S_\Lambda$. In the limit 
$\delta\Lambda\to 0$, the shell momentum $q$ reaches to $\Lambda$ 
from below. To make this limit well-defined, we should employ 
the left semi-open interval $\Lambda-\delta\Lambda<q\le\Lambda$. 
Hence we can say that the fluctuations with $q>\Lambda$ are 
incorporated in the Wilsonian effective action $S_\Lambda[\phi]$, 
while that with $q=\Lambda$ are not. It means that our infra-red cutoff 
$\theta(q-\Lambda)$ satisfies $\theta(0)=0$! 

\begin{figure}[htb]
\epsfxsize=0.9\textwidth
\centerline{\epsfbox{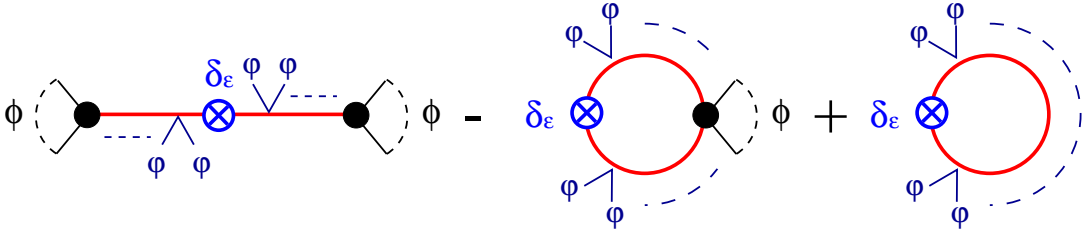}}
\caption{
These diagrams correspond to Eqs.\ (\ref{eq:dumbbell}) and (\ref{eq:ring}). 
The third graph has no dependence on the field $\phi(\Bq)$ but on the
VEV $\varphi$. 
It contributes to the LPA flow equation, and should be compared with the 
LPA Wegner-Houghton RGE.\protect\cite{rf:RGE} 
}
\label{fig:8}
\end{figure}

Since the `delta'-function $\delta_\varepsilon(q,\Lambda)$ lies on the 
$\Lambda$ derivative of the cutoff propagator, what we must check 
are the following two terms. One is a field $\phi(\Bq)$ dependent term,
\be
-\F{\bar\delta{S_\Lambda}}{\bar\delta\phi(\Bq)}\cdot
\left(\F{\D}{\D\Lambda}P_\Lambda\right)
\cdot\F{\bar\delta{S_\Lambda}}{\bar\delta\phi(-\Bq)}
+\F{\D}{\D\Lambda}P_\Lambda\cdot
\F{\bar\delta^2\hat S_\Lambda}{\bar\delta\phi(\Bq)
\bar\delta\phi(-\Bq)},
\ee
where the first term corresponds to the `dumbbell' diagram 
and the second term corresponds to the `ring' diagram in Fig.\ \ref{fig:8}. 
One can easily realize that the above equation is proportional to 
\be
\F{1}{q^2}\delta_\varepsilon(q,\Lambda)\left(\F{q^2/\Delta}{q^2/\Delta+
\Sigma_\Lambda^\varepsilon(q)}\right)^2
\stackrel{\varepsilon\to 0}{\longrightarrow}
\F{\delta(q-\Lambda)}{q^2+\Sigma_\Lambda^{\varepsilon=0}(q)},
\label{eq:dumbbell}
\ee
where we were taking account of the following relations,
\be
\F{\bar\delta{S_\Lambda^\varepsilon}}{\bar\delta\phi(\Bq)}\sim
\F{q^2}{q^2+\Delta\cdot\Sigma_\Lambda^{\varepsilon=0}}\cdot
\F{\bar\delta{S_\Lambda^{\varepsilon=0}}}{\bar\delta\phi(\Bq)}+
{\mbox{no significant terms}},
\ee
and 
\be
\F{\bar\delta^2\hat S_\Lambda^\varepsilon}{\bar\delta\phi(\Bq)
\bar\delta\phi(-\Bq)}\sim
\left(\F{q^2}{q^2+\Delta\cdot
\Sigma_\Lambda^{\varepsilon=0}(q)}\right)^2\cdot
\F{\bar\delta^2\hat S_\Lambda^{\varepsilon=0}}{\bar\delta\phi(\Bq)
\bar\delta\phi(-\Bq)}+{\mbox{no significant terms}},
\ee
for the functional derivative with respect to the field $\phi(\Bq)$. 

Another is the $\phi(\Bq)$ independent term which corresponds to 
the part evaluated in the Local Potential Approximation (LPA),
\be
-\F{\D}{\D\Lambda}P_\Lambda\cdot
S_\Lambda^{(2)}(q)=\F{1}{q^2}\delta_\varepsilon(q,\Lambda)
\cdot\F{q^2\Sigma_\Lambda^\varepsilon}
{q^2+\Delta\cdot\Sigma_\Lambda^\varepsilon}
\cdot (2\pi)^d\delta^d(\Bp+\Bq).
\ee
By taking $\varepsilon\to 0$, we find the sharp cutoff limit 
of this as
\be
\delta(q-\Lambda)\ln{\left(1+\Sigma_\Lambda^{\varepsilon=0}/q^2\right)}
\cdot (2\pi)^d\delta^d(\Bp+\Bq).
\label{eq:ring}
\ee
They correspond to the diagrams given in Fig.\ \ref{fig:8}. 

Using these results, the sharp cutoff limit of the Polchinski RG equation 
becomes, 
\begin{eqnarray}
&&\F{\D}{\D\Lambda}S_\Lambda^{\varepsilon=0}=
\F{1}{2}\int\F{d^d q}{(2\pi)^d}
\F{\delta(q-\Lambda)}{q^2+\Sigma_\Lambda^{\varepsilon=0}
(q,{\mbox{\small$\prod$}}
\varphi_i^{n_i})}\Bigg\{
\F{\bar\delta S_\Lambda^{\varepsilon=0}}{\bar\delta\phi(\Bq)}\cdot
\F{\bar\delta S_\Lambda^{\varepsilon=0}}{\bar\delta\phi(-\Bq)}
-\F{\bar\delta^2\hat{S}_\Lambda^{\varepsilon=0}}{\bar\delta\phi(\Bq)
\bar\delta\phi(-\Bq)}\Bigg\}\nonumber\\
&&\qquad\qquad\qquad
-\F{1}{2}\left(2\pi\right)^d\delta^d(0)
\int\F{d^d q}{(2\pi)^d}\delta(q-\Lambda)
\ln{\left(q^2+\Sigma_\Lambda^{\varepsilon=0}(q,{\mbox{\small$\prod$}}
\varphi_i
^{n_i})\right)}.
\label{eq:sharppol}
\end{eqnarray}
The canonical scaling of the momentum $p_\mu\D/\D p_\mu$ does not 
affect these results, since $\D\Delta/\D p_\mu\to 0$ as 
$\varepsilon\to 0$. 

\section{Comparison with the Wegner-Houghton equation}
\label{sec:5}

To confirm the equivalence between Eq.\ (\ref{eq:sharppol}) and the 
Wegner-Houghton 
RG equation, we substitute Eq.\ (\ref{eq:field}) to the Wegner-Houghton 
equation. 
Let us start from the following formula which gives the 
effective action $S_{\Lambda-\delta\Lambda}$ 
up to $O(\delta\Lambda^2)$. 
\be
S_{\Lambda-\delta\Lambda}=
S_\Lambda-\F{1}{2}
\F{\bar\delta S_\Lambda}{\bar\delta\phi_s} \Big |
\cdot\left(P_\Lambda^{-1}+\F{\bar\delta^2 S_\Lambda}{\bar\delta\phi_s
\bar\delta\phi_s}\Big |\right)^{-1}\cdot
\F{\bar\delta S_\Lambda}{\bar\delta\phi_s}\Big |
+\F{1}{2}{\rm Tr}\ln{\left(P_\Lambda^{-1}
+\F{\bar\delta^2 S_\Lambda}{\bar\delta\phi_s\bar\delta\phi_s}\Big |
\right)},
\label{eq:gauss11}
\ee
where $\phi_s$ denotes the `shell mode' whose support is given by the 
condition $p^2=\Lambda^2$. Dot ($\cdot$) denotes the integral 
$\int_{\Lambda-\delta\Lambda}^\Lambda d^dq/(2\pi)^d$. 
It can be realized that 
Eq.\ (\ref{eq:gauss11}) involves the higher contribution of 
$O(\delta\Lambda^2)$. First, we define $\hat{S}_\Lambda$ by,
\be
\F{\bar\delta^2 S_\Lambda}{\bar\delta\phi_s(\Bp)
\bar\delta\phi_s(\Bq)}\Big |_{\phi_s=0}\equiv
\Sigma_\Lambda(q,{\mbox{\small$\prod$}}\varphi_i^{n_i})
(2\pi)^d\delta^d
\left(\Bp+\Bq\right)+
\F{\bar\delta^2\hat S_\Lambda}{\bar\delta\phi_s(\Bp)
\bar\delta\phi_s(\Bq)}\Big |_{\phi_s=0}.
\label{eq:two2}
\ee
Here, $\Sigma_\Lambda$ is the same as $\Sigma_\Lambda^{\varepsilon=0}$ given
by Eq.\ (\ref{eq:2pfun}) before. 
The second term of the r.h.s. of Eq.\ (\ref{eq:two2}) 
is regular at $\Bq=-\Bp$. 

Let us rewrite Eq.\ (\ref{eq:gauss11}) in matrix notation. 
We define a matrix ${\bf M}$ by,
\be
{\bf M}_{p,q}\equiv
\F{\bar\delta^2\hat S_\Lambda }{\bar\delta\phi_s(\Bp)
\bar\delta\phi_s(\Bq)}\Big |_{\phi_s=0}.
\ee
The matrix ${\bf M}_{p,q}$ may have off-diagonal singularities, e.g. 
$\delta(\Bp+\Bq+{\bf k})$ due to $\varphi_i$. 
The first derivative of the effective action with respect to the field 
corresponds to a `vector' ${\bf v}$; 
\be
{\bf v}_q\equiv\F{\bar\delta S_\Lambda}{\bar\delta\phi_s(\Bq)} 
\Big |_{\phi_s=0}.
\ee

\begin{figure}[htb]
\epsfxsize=0.3\textwidth
\centerline{\epsfbox{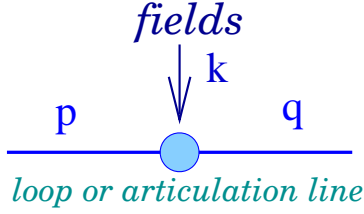}}
\caption{%
A diagram which external momenta $k$ flows in is drown.}
\label{fig:9}
\end{figure}

Using these notations, the r.h.s. of Eq.\ (\ref{eq:gauss11}) can be expressed 
by the following equation, 
\be
\F{1}{2}{\bf v}\TR\cdot
\left((P^{-1}+\Sigma){\bf 1}+{\bf M}\right)^{-1}
\cdot{\bf v}
-\F{1}{2}{\rm Tr}\ln\left((P^{-1}+\Sigma){\bf 1}+{\bf M}\right),
\ee
where the unit matrix ${\bf 1}$ corresponds to 
$(2\pi)^d\delta^d\left(\Bp+\Bq\right)$. 
We expand this with respect to the matrix ${\bf M}$. Since one momentum 
intergal $\int_{\Lambda-\delta\Lambda}^\Lambda d^dq$ brings a factor 
$\delta\Lambda$, only the first few terms can contribute to the RG flow 
equation, therefore
\be
\F{1}{2}{\bf v}\TR\cdot
\left((P^{-1}+\Sigma){\bf 1}\right)^{-1}
\cdot{\bf v}
-\F{1}{2}{\rm Tr}\ln\left((P^{-1}+\Sigma){\bf 1}\right)-
\F{1}{2}{\rm Tr}\left((P^{-1}+\Sigma){\bf 1}\right)^{-2}
\cdot{\bf M}+\cdots,
\ee
where dots $\cdots$ are the higher order in $\delta\Lambda$. 
One of the higher order contribution is written as follows,
\be
\F{\bar\delta S_\Lambda}{\bar\delta\phi_s}
\cdot P_s\cdot
\F{\bar\delta^2\hat S_\Lambda}{\bar\delta\phi_s\bar\delta\phi_s}
\cdot P_s\cdot
\F{\bar\delta S_\Lambda}{\bar\delta\phi_s}
\quad ,\label{eq:example}
\ee
where $P_s(q)$ is the propagator of the shell mode $\phi_s$ whose support is 
restricted to the region $\Lambda-\delta\Lambda<q\le\Lambda$. 
Eq.\ (\ref{eq:example}) corresponds to Fig.\ \ref{fig:9}. 

Since the cross section of the integral region of $p$ and that of $q$ is 
$O(\delta\Lambda^2)$, the contribution from the diagram given in 
Fig.\ \ref{fig:9} becomes 
$O( \delta\Lambda^2)$. 
If $k$ vanishes, the volume of the integral region above becomes 
the first order of $\delta\Lambda$, because two `spheres' completely coincide
with each other. 
\begin{figure}[htb]
\epsfxsize=0.4\textwidth
\centerline{\epsfbox{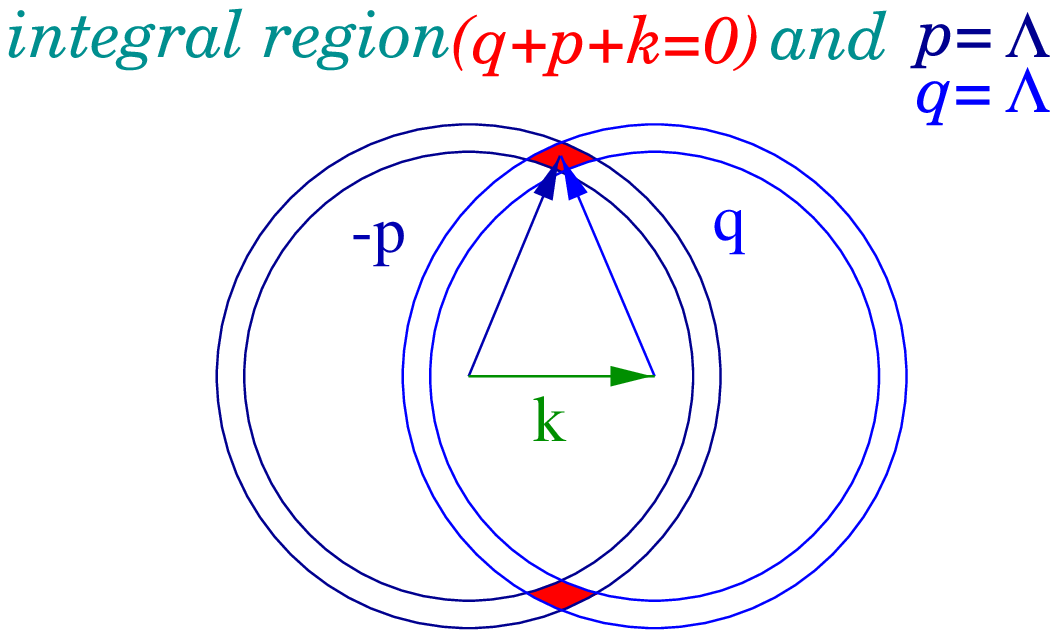}}
\caption{%
The integral region in the momentum space for the diagram given in 
Fig.\ \ref{fig:9}. The volume of the 
cross section of the two integral regions is order of $\delta\Lambda^2$. 
}
\label{fig:10}
\end{figure}
Hence the value of the RG beta-function jumps at the point $k=0$. 
However since the field $\phi$ is a smooth function of the momentum, 
not the distribution, we can drop it. It contributes through the 
distribution given in Eq.\ (\ref{eq:field}) by the combinations 
${\mbox{\small$\prod$}}\varphi_i^{n_i}$ 
with $\sum n_ip_i=0$. They are already taken in the beta-function by the 
${\mbox{\small$\prod$}}
\varphi_i^{n_i}$ dependence in the self energy $\Sigma_\Lambda$. 

Consequently, the Wegner-Houghton equation can be found as,
\begin{eqnarray}
&&S_\Lambda-S_{\Lambda-\delta\Lambda}=
\F{\delta\Lambda}{2}\int\F{d^d q}{(2\pi)^d}
\F{\delta(q-\Lambda)}{q^2+
\Sigma_\Lambda(q,{\mbox{\small$\prod$}}\varphi_i^{n_i})}\Bigg\{
\F{\bar\delta{S_\Lambda}}{\bar\delta\phi(\Bq)}\cdot
\F{\bar\delta{S_\Lambda}}{\bar\delta\phi(-\Bq)}
-\F{\bar\delta^2\hat S_\Lambda}{\bar\delta\phi(\Bq)
\bar\delta\phi(-\Bq)}\Bigg\}\nonumber\\
&&\qquad\qquad
-\F{\delta\Lambda}{2}\left(2\pi\right)^d\delta^d(0)
\int\F{d^d q}{(2\pi)^d}\delta(q-\Lambda)
\ln{\left(q^2+\Sigma_\Lambda(q,{\mbox{\small$\prod$}}\varphi_i^{n_i})\right)}
+O(\delta\Lambda^2).
\label{eq:whrt11}
\end{eqnarray}
This flow equation is nothing but the sharp cutoff limit of the Polchinski 
equation (\ref{eq:sharppol}).

\section{Conclusion and remarks}
\label{sec:6}

In this article, we investigated the cutoff scheme dependence of the 
Wilsonian effective action. It can be reinterpleted as the coordinate 
transformation on the theory space. It is written formally by 
Eq.\ (\ref{eq:trans}). We have studied it in two limiting cases. One is in 
the asymptotic region i.e. $t\to\infty$, and another is in 
the sharp cutoff limit i.e. $\varepsilon\to 0$. In the both cases, we 
could write down the cutoff scheme dependence so simple as to explore the 
RG flows. 

As we have shown in Sec.\ \ref{sec:3}, 
the scheme dependence of the renormalized trajectories in the 
asymptotic region $t\to\infty$ remains for the Wilsonian effective
action. Besides, the RG flow of the Wilsonian effective
action does not freeze in $t\to\infty$. 
The origin is as follows. The vertices of the Wilsonian 
effective action consist of two different quantities; the connected 
Green's function at high energy region ($p>\Lambda$) and the 1PI vertices at 
the low energy region ($p<\Lambda$). The boundary between these regions are 
connected scheme dependently. (See also Ref.\ \cite{rf:MOR1}.) 
Therefore {\it the Wilsonian effective action itself
is not a physical quantity}. 

Moreover, we have also shown the scheme independence of the 
Legendre effective action,\cite{rf:FE,rf:MOR1} 
or equivalently the 1PI building blocks of the 
Wilsonian effective action, on the renormalized trajectories. 
Recalling the statements in Sec.\ \ref{sec:3}, 
we can say that {\it if the RG flow of the dimensionful Legendre 
effective action 
$\Gamma_\Lambda[\phi]$ freezes on the renormalized trajectory in
the asymptotic region, i.e. in the statistical continuum 
limit $\Lambda_0\to\infty$, 
then our $\Gamma_\Lambda[\phi]$ should be 
scheme independent.} 

In the perturbation theory, it can be easily realized. Indeed, all 
the cutoff scheme dependent contributions, i.e. the coefficients of the 
divergences, are completely absorbed into certain counterterms 
order by order, and remaining finite terms are scheme independent 
in the limit $\Lambda_0\to\infty$. Of course, needless to say, 
we should insist on the common renormalization condition (or equivalently 
the subtruction rule). 
In the non-perturbative case, however, the problem will be more 
complicated, because 
the ordinary renormalization procedure will not work in general, 
e.g. for the theory around a non-Gaussian fixed point. 
Hence, the cancellation of divergences, and therefore the 
cutoff scheme dependent constants can be confirmed only 
case by case if possible. 

Turning to the Exact Renormalization Group, we can recapture it from 
another point of view. The scheme dependence of the 
counterterms corresponds to that of the fixed point and/or of the 
critical surface, and the scheme independence of the total solution 
can be appreciated by that of the renormalized trajectory in the 
asymptotic region. All these are described by a 
coordinate transformation (\ref{eq:trans}). For our purpose, 
it is sufficient to investigate the asymptotic region of Eq.\ (\ref{eq:trans}) 
without using the explicit solutions, since we have expected the 
asymptotic behavior $g_i(t)\sim g_i^R e^{d_it}$ and do not need the 
explicit value $g_i^R$. Once we assume existence of the 
asymptotic region, then the scheme independence of 
$\Gamma_\Lambda[\phi]$ as $\Lambda\to 0$ is confirmed. 
(Recall the discussion in Sec.\ \ref{sec:3}.) 
For massive theories, the scheme 
dependence of $\Gamma_\Lambda[\phi]$ decays like 
$\exp(-dt)\sim(\Lambda/M_R)^{d}$ as $t\to\infty$. 
Instead, for the massless case, one may start from a massive case and then 
letting $M_R\to 0$. 

We also confirm the equivalence between the Wegner-Houghton equation and 
the Polchinski equation in the sharp cutoff limit. 
It seems that these equations 
are much different from each other even though they are expected to be 
equivalent. 
We can prove equivalence of these two ERGs by help of 
Eq.\ (\ref{eq:trans}) which describes the 
scheme dependence of the Wilsonian effective action. 
The superficial difference occurs by 
the strong scheme dependence of the Wilsonian effective action. As we showed, 
the crucial cutoff scheme dependence of the Wilsonian effective action 
lies in the external legs. 

Finally, we would like to comment on the cutoff scheme dependence of 
the approximate solutions. 
The ERG flow equations are approximated by projecting them onto 
smaller dimensional subspaces of the original theory space. 
In the derivative expansion, for example, we may employ these 
subspaces as the space of a finite number of the coefficient 
functions $\{V_0,V_2,\cdots,V_k^i\}$ defined by the 
following equation, 
\be
S_\Lambda[\phi]=\int d^dx\left\{
V_0(\phi)+\F{1}{2}(\D_\mu\phi)^2V_2(\phi)
+\F{1}{2}(\Box\phi)^2 V_4^1(\phi)+\cdots\right\},
\label{eq:expans}
\ee
The subscript $k$ 
of the coefficient function denotes the degrees of the derivatives and 
the superscript $i$ of it labels the independent $k$-th derivative vertices. 
Then, the ERG flow equations are reduced to the coupled partial differential
equations for the coefficient functions $V_k^i(\phi)$. 
One can easily improve the approximation systematically by enlarging 
the subspace $\{V_0,V_2,\cdots,V_k^i\}$ step by step. 
Especially, the approximation with $k=0$ is called the `local potential 
approximation' (LPA). 
This procedure preserves the non-perturbative nature of the ERG flow 
equations. 

The scheme dependence given by Eq.\ (\ref{eq:delta1}) 
are infinitely enhanced in the derivative expansion. 
By dimensional analysis, the Taylor expansion of $\delta P(q)$,
whose value changes rapidly near the infra-red cutoff $q\sim\Lambda$,
is the expansion with respect to the combination $q^2/\Lambda^2\gg 1$. 
Therefore the scheme dependence of 
the coefficient functions $V_i^k(\phi)$ in Eq.\ (\ref{eq:expans}) 
become stronger and stronger as the infra-red cutoff $\Lambda$ decreases. 
The scheme dependence of $V_k^i(\phi)$ behaves as $1/\Lambda^{2k}$. 
At last it diverges in the limit $\Lambda\to 0$. It means that the 
derivative expansion and the limit $\Lambda\to 0$ do not commute. 
Hence the cutoff scheme dependence of $V_i^k(\phi)$ is strong enough to 
prevent the physical predictions. The physical information is completely 
washed off except for the potential part $V_0(\phi)$ which is the
1PI effective potential. 

The RG beta-functionals of the coefficient functions of 
$\Gamma_\Lambda[\phi]$ like Eq.\ (\ref{eq:expans}) 
are also cutoff scheme dependent in the region $t\to\infty$, since 
by the dimensional analysis, the expanding parameter there is 
$\D/\varepsilon\sim\D/\Lambda$ where 
$\varepsilon$ stands for the smoothness parameter given in 
Sec.\ \ref{sec:4}. 
The higher derivative contributions finally 
overcome the suppression factor $(\Lambda/M_R)^{d}$. 
Hence, the RG beta-functionals 
of the higher derivative operators blow up to infinity. 
In the limit $t\to\infty$, the cutoff scheme approaches towards to 
the sharp one, 
since $\varepsilon\sim\Lambda\to 0$. It is known 
that these diverging series can be summed up and lead to 
non-analytical momentum dependence of the vertices, 
e.g. $\sqrt{p_\mu p_\mu}$. This spurious scheme dependence can be 
avoided if we work on the sharp cutoff Legendre flow 
equation \cite{rf:MOR1,rf:MOR4}. 

\section*{Appendix} 

Equation (\ref{eq:connect}) in Sec.\ \ref{sec:3} is driven as follows. 
First, let us consider the positive definite deviation of the cutoff 
propagator $\delta P(q)=\delta P_+(q)> 0$, since we call for the Gaussian 
integral with positive $\delta P(q)$. 
Then one can find
\begin{eqnarray}
&&\exp{\left( \F{1}{2}\delta D_+[\delta/\delta\phi] \right)}
\exp{\left( -S_t\left[ \phi \right] \right)}=
\exp{\left( \F{1}{2}\delta D_+[\delta/\delta\phi] \right)}
\exp{\left( -S_t\left[ \phi \right] +J\cdot\phi\right)}
\Big|_{J=0}
\nonumber\\
&&\quad=
\exp{\left( \F{1}{2}\delta D_+[\delta/\delta\phi] \right)}
\exp{\left( -S_t\left[ \delta/\delta J\right] \right)}
\;{\rm e}^{J\cdot\phi}
\Big|_{J=0}
\nonumber\\
&&\quad=
\exp{\left( -S_t\left[ \delta/\delta J\right] \right)}
\exp{\left( \F{1}{2}\delta D_+[ J ]+J\cdot\phi \right)}
\Big|_{J=0}
\nonumber\\
&&\quad\propto
\exp{\left( -S_t\left[ \delta/\delta J\right] \right)}
\int D\phi'
\exp{\left(
-\F{1}{2}\phi'\cdot\delta P_+^{-1}\cdot\phi'
+J\cdot(\phi+\phi')\right)}\Big|_{J=0}
\nonumber\\
&&\quad=
\int D\phi'
\exp{\left(
-\F{1}{2}\phi'\cdot\delta P_+^{-1}\cdot\phi'
-S_t\left[ \phi+\phi'\right]\right)}
\nonumber\\
&&\quad=
\exp{\left(-\F{1}{2}\phi\cdot\delta P_+^{-1}\cdot\phi
\right)}
\int D\phi'\exp{\left(
-\F{1}{2}\phi'\cdot\delta P_+^{-1}\cdot\phi'
-S_t\left[ \phi'\right]
+\phi'\cdot\delta P_+^{-1}\cdot\phi\right)}\nonumber\\
&&\quad=
\exp{\left(-\F{1}{2}\phi\cdot\delta P_+^{-1}\cdot\phi
+W_t[J=\delta P_+^{-1}\cdot\phi]\right)},
\label{eq:AA}
\end{eqnarray}
where $\delta D_+[J]$ is given by, 
\be
\delta D_+[J]=\int\F{d^d q}{(2\pi)^d}
J(q)\delta P_+(q)J(-q).
\ee
The negative part of the deviation $\delta P_-(q)$ needs the special care,
since the path-integral in the fourth line does not converge. 
However our final result can be hold also for the negative part 
$\delta P_-(q)$, 
since Eq.\ (\ref{eq:connect}) is the identity of $\delta P$. In other words, 
Eq.\ (\ref{eq:connect}) means the graph by graph correspondence of the 
Feynman diagrams. It does not restrict our observations in 
Secs.~\ref{sec:3}-\ref{sec:5}, since we need the diagramatical 
representation of Eq.\ (\ref{eq:AA}).


\begin{thebibliography}{99}
\bibitem{rf:RGE} K.G.~Wilson and I.G.~Kogut, \JL{Phys.~Rep.,12,1974,75}.\\
F.J.~Wegner and A.~Houghton, \PR{A8,1973,401}\\
J. Polchinski, \NP{B231,1984,269}.\\
G.~Keller, C.~Kopper and M.~Salmhofer, \JL{Helv.~Phys.~Acta,65,1992,32}.
\bibitem{rf:chiral} 
U.~Ellwanger and C.~Wetterich, \NP{B423,1994,137}.\\
D.U.~Jungnickel and C.~Wetterich, Lectures given at Workshop 
on the Exact Renormalization Group, Faro, Portugal, 10-12 Sept. 
1998, hep-th/9902316; \PR{D53,1996,5142}; 
\JL{Eur.~Phys.~J.,C1,1998,669}; \andvol{C2,1998,557}; 
\PL{B389,1996,600}; 
Heidelberg preprints HD-THEP-96-40, hep-ph/9610336.\\
J.~Berges, D.U.~Jungnickel, and C.~Wetterich, 
\PR{D59,1999,34010}.\\
M.~Reuter and C.~Wetterich, \PR{D56,1997,7893}.\\
K-I.~Aoki, K.~Morikawa, J-I.~Sumi, H.~Terao and M.~Tomoyose, 
\PTP{97,1997,479}; \PTP{102,1999,1151}; 
hep-th/9908043 to be published in Phys.~Rev. {\bf D61}.\\
H.~Kodama and J-I.~Sumi, hep-th/9912215, to appear in
Prog. Theor. Phys.\\
K.~Kubota and H.~Terao, \PTP{102,199,1163}.\\
K-I.~Aoki, K.~Takagi, H.~Terao and M.~Tomoyose, hep-th/0002038.
\bibitem{rf:MOR2} 
C.~Wetterich, \JL{Z.~Phys.,C57,1993,451}. \\
N.~Tetradis and C.~Wetterich, \NP{B422,1994,541}. \\
T.R.~Morris, \PL{B329,1994,241}. 
\bibitem{rf:MOR1} 
T.R. Morris, \IJMP{A9,1994,2411}.
\bibitem{rf:MOR4} 
T.R.~Morris, \NP{B495~[FS],1997,477}.
\bibitem{rf:SO} 
K-I.~Aoki, K.~Morikawa, W.~Souma, J-I.~Sumi and H.~Terao, 
\PTP{95,1996,409}; \PTP{99,1998,451}
\bibitem{rf:FE} C.~Wetterich, \PL{B301,1993,90}.\\
M.~Bonini, M.~D'Attanasio, and G.~Marchesini, \NP{B409,1993,441}.
\bibitem{rf:CC} 
J.F.~Nicol, T.S.~Chang and H.E.~Stanley, \PRL{33,1974,540}.\\
T.R.~Morris, \PL{B334,1994,355}.
\bibitem{rf:MOR3} 
T.R.~Morris, \PL{B334,1994,355}.
\bibitem{rf:lpa}
A.~Hazenfratz and P.~Hazenfratz, \NP{B270~[FS16],1986,269}. 
\bibitem{rf:ALF} 
M.~Alford, \PL{B336,1994,237}.\\
N.~Tetradis and C.~Wetterich, \NP{B422,1994,541}.
\bibitem{rf:MO} 
K-I.~Aoki, K.~Morikawa, W.~Souma, J-I.~Sumi and H.~Terao, 
\PTP{99,1998,451}.
\bibitem{rf:BHLM} 
R.D.~Ball, P.E.~Haagensen, J.I.~Latorre and E.~Moreno, \PL{B347,1995,80}.
\bibitem{rf:WEIN}
S.~Weinberg, ``{\it Ultraviolet divergences in quantum theories of 
gravitation}'', (Cambridge Univ. Press., Cambridge, Eng., 1979) pp.790.
\end{thebibliography}
\end{document}